\documentclass[amsmath,amssymb,aps,letterpaper,prb,twocolumn,longbibliography,superscriptaddress]{revtex4-2}
\usepackage[english]{babel}
\usepackage[utf8]{inputenc}
\usepackage[T1]{fontenc}
\usepackage{indentfirst}
\usepackage{amsmath}
\usepackage{amssymb}
\usepackage{amsthm}
\usepackage{proof}
\usepackage{eufrak}
\usepackage{graphicx}
\usepackage{psfrag}

\allowhyphens

\newcommand{\VV}{\mathcal{V}}

\newcommand{\Pe}{\mathcal{P}}

\newcommand{\UU}{\mathcal{U}}
\newcommand{\field}{\mathbb{Z}}
\newcommand{\complex}{\mathbb{C}}
\newcommand{\egesz}{\mathbb{Z}}

\newcommand{\valos}{\mathbb{R}}

\newcommand{\ordo}{\mathcal{O}}

\newcommand{\vev}[1]{\left\langle #1 \right\rangle}
\newcommand{\ket}[1]{{\left|#1\right\rangle}}
\newcommand{\bra}[1]{{\left\langle #1\right|}}

\newcommand{\R}{{\mathbb R}}

\newcommand{\Z}{{\mathbb Z}}
\newcommand{\C}{{\mathbb C}}

\newcommand{\F}{{\mathbb F}}

\newtheorem{prop}{Proposition}[section]

\newtheorem{cor}{Corollary}[section]


\usepackage{ifpdf}

\ifpdf
\usepackage{epstopdf}
\usepackage[pdftex,colorlinks,urlcolor=blue,citecolor=blue,linkcolor=blue]{hyperref}
\else
\usepackage[hypertex,colorlinks,urlcolor=blue,citecolor=blue,linkcolor=blue]{hyperref}
\fi
\begin{document}

\title{
  Construction and the ergodicity properties of dual unitary quantum circuits
  }

\author{M\'arton Borsi}
\affiliation{MTA-ELTE “Momentum” Integrable Quantum Dynamics Research Group, Department of Theoretical Physics, Eötvös
  Loránd University, P\'azm\'any P\'eter stny. 1A, Budapest 1117, Hungary}
\author{Bal\'azs Pozsgay}
\affiliation{MTA-ELTE “Momentum” Integrable Quantum Dynamics Research Group, Department of Theoretical Physics, Eötvös
  Loránd University, P\'azm\'any P\'eter stny. 1A, Budapest 1117, Hungary}

\begin{abstract}
We consider one dimensional quantum circuits of the brickwork type, where the fundamental quantum gate is dual
unitary. Such models are solvable: the dynamical correlation functions of the infinite temperature ensemble can be
computed exactly. We review various existing constructions for dual unitary gates and we supplement them
with new ideas in a number of cases. We discuss connections with various topics in physics and mathematics, including
quantum information theory, tensor networks for the AdS/CFT correspondence (holographic error correcting codes),
classical combinatorial designs (orthogonal Latin squares), planar algebras, and Yang-Baxter maps.
Afterwards we consider the ergodicity properties of a special class of dual
unitary models, where the local gate is a permutation matrix. 
We find an unexpected phenomenon: non-ergodic behaviour can manifest itself in multi-site
correlations, even in those cases when the one-site correlation functions are fully chaotic (completely
thermalizing). We also discuss the circuits built out of perfect tensors. They appear locally as the most chaotic and
most scrambling circuits, nevertheless they can show global 
signs of non-ergodicity: if the perfect tensor is constructed from a linear map over finite fields, then
the resulting circuit can show exact quantum revivals at unexpectedly short times. A brief mathematical treatment
of the recurrence time in such models is presented in the Appendix (written by Roland Bacher and Denis Serre).
\end{abstract}

\maketitle

\section{Introduction}

Interacting many body systems are typically not amenable to analytic solution, neither in classical nor in quantum
mechanics, and their exact treatment is possible only in a very limited set of models.
Somewhat trivial examples are models which are equivalent to free
bosons of free fermions: in these cases the dynamics is one-body reducible and all correlation functions can be computed
using Wick's theorem. 
If there are interactions in the system, then one needs to impose some sort of restrictions on the dynamics such that it
remains tractable.

One class of such systems are the so-called one dimensional integrable models.
In this case
solvability is guaranteed by the existence of a large
(typically infinite) family of local conserved charges, leading to completely elastic and factorized scattering of
the quasi-particles \cite{caux-integrability}. This implies the lack of ergodicity, and as a result isolated integrable
models equilibrate 
to Generalized Gibbs Ensembles (see the review \cite{essler-fagotti-quench-review}).

Dual unitary (DU) quantum circuits are an other class of exactly solvable systems
\cite{dual-unitary-1,dual-unitary-2,dual-unitary-3}.  In these models the solvability follows from a special property of
the basic two-site quantum gate: dual unitarity means that the fundamental quantum  gate is unitary also when viewed as
a generator of dynamics in the space direction. The dual unitarity condition is equivalent to having a two-site unitary
gate with maximal bipartite operator entanglement \cite{dual-unitary--bernoulli}. The distinguishing physical behaviour
of DU circuits is that the dynamical two-point correlation functions (computed in the infinite temperature ensemble) are non-zero
only along light cones \cite{dual-unitary-3}. DU circuits can be chaotic or integrable (precise definitions will be
given in the main text); it is remarkable that even the chaotic DU circuits are solvable.

DU circuits received considerable attention in the recent years, and their correlation
functions and entangling properties were studied in depth, see for example
\cite{sarang-lamacraft-kicked-DU,dual-gliders,dual-gliders-2,dual-unitary-4,dual-kicked,max-velocity,lorenzo-bruno,dual-unitary--bernoulli,entanglement-barrier}. Their
computational power from an information theoretical perspective was considered in \cite{dual-computational}. A full
classification of DU gates is available for local dimension $N=2$ \cite{dual-unitary-3}, but some constructions are known in the
literature also for $N\ge 3$ which will be reviewed in Section \ref{sec:constr}. A numerical method for the production
of DU gates was presented in \cite{dual-ensembles-1}.

A special class of DU gates are those which have maximal entangling
power: these are perfect tensors \cite{AME-1,AMEcomb1,chaos-qchann}. In quantum information theory, a perfect tensor is a
vector of a multi-site Hilbert space, which has maximal bipartite entanglement for all possible partitioning of its
sites (also called parties or tensor legs). Alternatively, for an even number of legs they can be seen as multi-unitary operators 
for every bi-partitioning of the legs into two subsets with equal size \cite{AMEcomb2}.
Such states are also called Absolutely Maximally Entangled
(AME) states, and they have been studied in a large number of works, see for example \cite{AMEcomb1,four-AME,AMEcomb2,AMEcomb3}
and references therein. A list of AME states with various numbers of legs and local dimensions is available at the homepage
\cite{AME-list}.  Dual unitary gates which are also perfect tensors correspond
to AME states with four legs. It is known that such 
states exist for all local dimensions except for $N=2$ \cite{AME-bounds}; the last remaining open case of $N=6$ was
solved very recently \cite{euler36}. Perfect tensors can be used as quantum error correcting codes, and they are
essential ingredients in the 
tensor network models of the AdS/CFT bulk-boundary correspondence (see the seminal paper \cite{ads-code-1} and the
recent review \cite{holocode-review}). 

Dual unitary gates can be seen as ``broken'' or ``imperfect'' versions of the perfect tensors with four legs.
Research on quantum error correcting codes and especially on holographic codes led to the introduction of such
``broken'' or weaker
versions of ``perfectness''. The work \cite{perfect-tangles} introduced perfect tangles, which are multi-leg tensors that
have maximal entanglement for all ``planar'' bi-partitions; the dual unitary gates correspond to perfect tangles with
four legs. The same definition was given independently in \cite{block-perfect-tensor} and in \cite{planar-AME}, where the same
objects were called ``block perfect tensors'' and ``planar maximally entangled states'', respectively.
The tri-unitary quantum gates considered recently in \cite{triunitary} are six-leg tensors which are
perfect tangles.

Time evolution in a quantum circuit with a perfect tensor was studied  already in 2015 in \cite{chaos-qchann}. It was shown here
that the resulting circuit generates ballistic growth of operator entanglement, and  it is a good toy model for
scrambling of quantum information. More recently circuits built from so-called perfect DU gates were also studied in
\cite{dual-unitary--bernoulli}, where they were called ``Bernoulli circuits'' (for a definition see the main
text). These circuits can be seen as the most 
chaotic ones, at least from a local perspective: all dynamical correlation functions vanish in the infinite temperature
ensemble. In the tensor network models of AdS/CFT the analogous behaviour was observed in \cite{ads-code-perturb}.
Nevertheless, such 
circuits can also show global signs of integrability \cite{chaos-qchann}; we discuss this phenomenon in Section \ref{sec:perfect}.

Dual unitary matrices have a connection with pure mathematics, which has not yet been noticed in the relevant condensed
matter 
literature: they appear in the study of planar algebras.
This theory originates from the work of Vaughan Jones
\cite{jones-planar}, building on earlier works on von Neumann algebras (see the book \cite{subfactors-book} of Jones and
Sunder). The notion of dual unitarity (called bi-unitarity in the mathematical literature) appears
 in this theory. Implications for quantum information theory were studied in \cite{biunitary-qinf1}  and
more recently in \cite{biunitary-qinf2}. The general concept of bi-unitarity treated in these works accommodates not
only the actual DU gates, but also complex Hadamard matrices, quantum error correcting codes, and quantum Latin squares.
The very special case of dual unitary permutation matrices was studied already in
1996 in \cite{biunitary-permutations}, where a complete enumeration was given for local dimensions $N=2$ and $N=3$.

In this work we set ourselves two goals:
\begin{enumerate}
\item We collect a number of constructions for DU gates that appeared in
various places in the literature, and we complement them with new ideas. After a technical
introduction in Section \ref{sec:du} the constructions are discussed in Section \ref{sec:constr}.
\item We treat the ergodicity properties of DU circuits with permutation maps. 
These circuits can be understood 
as classical cellular automata, but it is useful to treat them also as quantum objects. We 
 uncover interesting physical phenomena in these models. We show that there are models which appear fully chaotic on the
 level of one-site correlations, but which display 
 non-ergodicity if we look at the multi-site reduced density matrices. The possible existence of such behaviour was
 already mentioned in \cite{dual-unitary-3} (see also the so-called ``still gliders'' found in \cite{thermalizing-boundaries})
 and now we find concrete examples for this; this discussion is presented in
 Section \ref{sec:dupm}. 
Afterwards we also investigate global signs of non-ergodicity in DU circuits built from perfect maps. These findings are
presented in Section \ref{sec:perfect}, complementing the relevant results of the earlier work \cite{chaos-qchann}. A
mathematical treatment of the recurrence time in a special class of models (given by linear maps over finite fields) is
given in Appendix \ref{sec:app}.
\end{enumerate}

 \section{Dual unitary circuits}

\label{sec:du}

We consider a spin chain with a Hilbert space $\mathcal{H}=\otimes_{j=1}^L \complex^N$, where the local dimension $N$ is
a fixed number of the system, and $L$ is the volume which we assume to be an even number. We build a quantum cellular
automaton on this Hilbert space, such that the time evolution is dictated by a quantum circuit of the brickwork
type. The fundamental object is a two site unitary gate $U$, which is a linear operator on $\complex^N\otimes \complex^N$. If
$\ket{\Psi(t)}\in \mathcal{H}$ is the state of the system at time $t$, then the
update rule of the QCA is such that
\begin{equation}
  \label{eom}
  \ket{\Psi(t+1)}=
  \begin{cases}
    \VV_1\ket{\Psi(t)} & \text{ for } t=2k+1\\
      \VV_2\ket{\Psi(t)} & \text{ for } t=2k,\\   
  \end{cases}
 \end{equation}
where the alternating update operations are
\begin{align}
&  \VV_1=U_{12}U_{34}\dots U_{L-1,L}\\
&  \VV_2=U_{23}U_{45}\dots U_{L,1}.
\end{align}
Here periodic boundary conditions are assumed. For further use we also introduce the Floquet operator
\begin{equation}
  \label{VVdef0}
 \VV=\VV_2\VV_1,
\end{equation}
which generates time evolution from time $t$ to $t+2$.

We are interested in the special case when $U$ is a dual-unitary operator
\cite{dual-unitary-1,dual-unitary-2,dual-unitary-3}. The dual unitary condition is formulated as follows.

\begin{figure}[t]
  \centering
\includegraphics{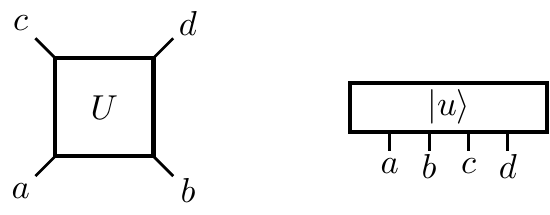}
      \caption{Pictorial representation of a DU gate. On the left the gate is pictured as a four leg tensor, which can
        generate unitary dynamics in both the time and space directions. For forward time evolution the incoming indices
        are $a$ and $b$, whereas for ``forward space evolution'' they are $c$ and $a$. On the right the same object is
        pictured as a vector of a fourfold product space. If the gate is DU, then the vector $\ket{u}$ is maximally
        entangled with respect to subsystems given by the pairs $(a,b)$, $(a,c)$ and their complements. It is a
        perfect tensor if maximal entanglement also holds for the subsystem given by the pair $(a,d)$.} 
  \label{fig:du}
\end{figure}

Let us choose
a basis in $\complex^N$ and denote $U^{cd}_{ab}$ the matrix elements of the operator $U$ in the given basis. We use the
reshuffled matrix $U^R$ given by
\begin{equation}
  (U^R)^{db}_{ca}=U^{cd}_{ab},
\end{equation}
and say that $U$ is dual unitary if it is unitary and
also $U^R$ is unitary. For a pictorial representation see Fig. \ref{fig:du}.

Using the SWAP gate $\Pe$ given by
\begin{equation}
  \Pe^{cd}_{ab}=\delta_{ad}\delta_{bc}
\end{equation}
we can see that $\Pe U^R=U^{t_2}$, where the superscript $t_2$ stands
for partial transpose in the second component. The swap gate is unitary, thus a bipartite unitary matrix $U$ is dual unitary, iff its
partial transposes are unitary. The latter property was called ``bi-unitary'' in the earlier mathematical literature
dealing with von Neumann algebras \cite{subfactors-book,biunitary-permutations}. The simplest dual unitary gate is the
SWAP gate itself.  

The dual unitarity condition implies a certain space-time duality: if we picture the resulting time evolution as a
tensor network, then this network describes unitary evolution also in the space direction. This property has important
consequences which we discuss below in Section \ref{sec:corr}. 

A special class of DU matrices are the so-called perfect tensors (or 2-unitaries) \cite{AMEcomb2}.
In order to introduce them we define one more reshuffling as
\begin{equation}
  (U^D)^{bc}_{ad}=U_{ab}^{cd}.
\end{equation}
It can be seen that $U^D$ acts ``from diagonal to diagonal'', see also Fig. \ref{fig:du}.
A DU gate is a perfect tensor if $U^D$ is also unitary matrix.
In this work we refer to such matrices as perfect DU gates.
The algebraic variety of such matrices was studied  in \cite{bipartite-unitaries}. The resulting quantum
circuits were analyzed in \cite{dual-unitary--bernoulli}, where they were called Bernoulli gates. It is known that
perfect tensors exist for all integers $N\ge 3$: the long standing open problem of the case $N=6$ was solved very recently
\cite{euler36}. 

For qubits DU gates were completely classified in \cite{dual-unitary-3}, but there is no general
parametrization of DU gates known for $N\ge 3$. Instead, several concrete constructions appeared in the literature. We
discuss these 
constructions later in Section \ref{sec:constr}.

\subsection{Correlation functions and conserved charges}

\label{sec:corr}

Physical properties of dual unitary circuits were studied in depth in the literature \cite{sarang-lamacraft-kicked-DU,dual-gliders,dual-gliders-2,dual-unitary-4,dual-kicked,max-velocity,dual-unitary--bernoulli,entanglement-barrier}. One of the central results is
that one-site correlation functions are non-zero only along light cones. To be more precise, let us consider two one-site
operators $o_1(x)$ and $o_2(y)$ and the infinite temperature average of the dynamical correlation function
\begin{equation}
\vev{o_1(x,t)o_2(y,0)}\equiv \frac{1}{N^L}
  \text{Tr} \Big(o_1(x,t)o_2(y,0)\Big),
\end{equation}
where it is understood that time evolution of the operator is dictated by the adjoin action of the Floquet operator
$\VV$. The central result of \cite{dual-unitary-3} is that such two-point functions can be non-zero only if
$|x-y|=t$. Furthermore, their precise values are determined by the eigenvalues of two quantum channels (light cone
propagators for the operators). To be precise, let us define the linear maps $M_\pm$ acting on one-site operators as
\begin{equation}
  \label{Mpm}
  \begin{split}
    M_+(o)&=\frac{1}{N}\text{Tr}_1 \Big(U^\dagger (o\otimes 1)U\Big)\\
     M_-(o)&=\frac{1}{N}\text{Tr}_2 \Big(U^\dagger (1\otimes o)U\Big).\\
  \end{split}
\end{equation}
Eigenvalues of these maps determine the behaviour of the correlation functions along the right moving and left
moving light cones, respectively. An eigenvalue $\lambda=1$ corresponds to a conserved one site operator, which we also
call glider (see below). Note that the identity operator $o=1$ always has $\lambda=1$ by unitarity.
Eigenvalues $|\lambda|=1$ different from 1 describe oscillating behaviour,
and eigenvalues $|\lambda|<1$ correspond to thermalizing modes. If all non-trivial eigenvalues are thermalizing, then
the circuit was interpreted as ergodic and mixing \cite{dual-unitary-3}.

A dual unitary model can also have multi-site conserved charges.
In order to discuss them, we introduce the concept of gliders.
Following \cite{clifford-qca-gliders} we say that a short range operator
$\ordo(x)$ positioned at site $x$ is a glider 
if its time evolution gives
\begin{equation}
  \ordo(x,t+1)=\ordo(x\pm 1,t),
\end{equation}
where the signs $+$ and $-$ stand for right moving and left moving gliders, respectively. In the formula above time
evolution is understood in the Heisenberg picture, dictated by the e.o.m. \eqref{eom}.
The equation above means that gliders do not suffer operator spreading, and summation
over $x$ gives an extensive conserved charge of the model.

It was shown in  \cite{dual-gliders-2} that in a DU circuit all conserved charges come from gliders.

It is useful to extend the definition of the gliders to allow for periodic oscillations, so that
\begin{equation}
  \label{glider1}
  \ordo(x,t+1)=e^{i\phi}\ordo(x\pm 1,t),
\end{equation}
where $e^{i2n \phi}=1$ with some $n$. In such cases the operator appears exactly conserved if we sample the system
stroboscopically with period $2n$.

The relation \eqref{glider1} can be written more explicitly as the exchange relation
\begin{equation}
  \label{glider2}
\ordo(x) \UU^{-2}\VV =e^{2i\phi} \UU^{-2}\VV  \ordo(x),
\end{equation}
where $\UU$ is the cyclic shift operator on the lattice and $\VV$ is the Floquet operator defined in \eqref{VVdef0}. 
For left
moving gliders we have an analogous relation  with $\UU^{-2}$ exchanged by $\UU^2$. 

The fact that all conserved charges are gliders makes the
search for them especially simple: One has to look for solutions of the linear equation \eqref{glider2} within the
vector space of local operators, with some
eigenvalues $\lambda$  that are roots of unity.

Equation \eqref{glider2} implies that gliders form an operator algebra which is closed under addition and
multiplication. Therefore, if a DU 
circuit has at least one conserved charge, then it has infinitely many: equal-time products of
gliders moving in the same direction are also gliders \cite{dual-gliders,dual-gliders-2}.

As an alternative to solving \eqref{glider2}, we can also consider the multi-site transfer matrices that
propagate extended local operators 
along light cone directions \cite{dual-unitary-3} and look again for the number of eigenvalues that are roots of unity.  These
transfer matrices were first introduced in \cite{lamacraft-circ,thermalizing-boundaries} and they are constructed as follows.

For a DU gate $U$ we introduce two more two-site gates as
\begin{equation}
  V=\Pe U^\dagger,\qquad V'=\Pe U^R.
\end{equation}
The action of the linear map in $M_+$ from \eqref{Mpm} can be rewritten as the action of a transfer matrix on the
doubled Hilbert space $\complex^N\otimes \complex^N$. Let $A$ stand for an auxiliary space and define the Matrix Product
Operator 
\begin{equation}
  t=\frac{1}{N}\text{Tr}_A\left[ V'_{2,A}V_{1,A}\right].
\end{equation}
In this case the eigenvectors of $t$ correspond to the eigen-operators of $M_+$. The extension to multi-site
channels is a transfer matrix acting on an auxiliary chain of length $2\alpha$ such that
\begin{equation}
  \label{talpha}
   t_\alpha=\frac{1}{N}\text{Tr}_A\left[V'_{2\alpha,A}\dots V'_{\alpha+1,A}V_{\alpha,A}\dots V_{2,A}V_{1,A}\right].
\end{equation}
This transfer matrix also appeared in \cite{lamacraft-circ}, see eq. (7) there. It describes the evolution of multi-site
operators along the positive light directions, for a maximal range of $2\alpha-1$ (for a more detailed derivation see
\cite{lamacraft-circ}).  The number of right moving gliders of maximal range $2\alpha-1$ can be found simply by counting
the eigenvalues $\lambda$ of the operator $t_\alpha=1$ which are roots of unity. For the left moving gliders one just performs
a space reflection of the original gate $U$.

In DU circuits with perfect tensors a very special phenomenon happens:
It was shown in \cite{dual-unitary--bernoulli} that in such cases all non-trivial non-equal-time correlations
are zero in the infinite 
temperature ensemble. This means that the spectra of the $t_\alpha$ (and their space reflected variants) are trivial:
they have a single trivial eigenvalue $\lambda=1$ and all remaining eigenvalues are zero.
This forbids the existence of local conserved charges in such models. Perfect tensors
should be interpreted as ``most ergodic'' in the 
hierarchy of DU circuits. The vanishing of the connected correlation functions in tensor network models with perfect
tensors was also observed in \cite{ads-code-perturb}. The property of ``maximal ergodicity'' underlies the application
of perfect tensors in lattice models of the AdS/CFT correspondence \cite{ads-code-1,holocode-review}. 

\subsection{Entanglement of operators and states}

It is useful to discuss the entanglement properties of dual unitary gates. The standard method is to picture the
operator as a vector of a doubled Hilbert space (see Fig. \ref{fig:du}).

To every unitary operator $U$ acting on a
bipartite system we can associate a four partite
state $\ket{u}\in \mathcal{H}_1\otimes\mathcal{H}_2\otimes\mathcal{H}_3\otimes\mathcal{H}_4$, where each
$\mathcal{H}_j\simeq \complex^N$ and in a concrete basis the components are $u_{abcd}=U_{ab}^{cd}/N$. The factor $1/N$
is added so that $\ket{u}$ is normalized to unity. Entanglement properties of the gate $U$ can be understood as
different sorts of entanglement of the state $\ket{u}$.

Let us consider reduced density matrices of the states $\ket{u}$ corresponding to pairs of sites. To this order we
introduce the set $S=\{1,2,3,4\}$ of ``sites'' or ``parties'' and define the reduced density matrices
\begin{equation}
\rho_{\alpha\beta}=\text{Tr}_{S\setminus \{\alpha,\beta\}} \ket{u}\bra{u},\quad \alpha,\beta\in S.
\end{equation}
The unitarity of the gate $U$ implies that $\rho_{12}$ and $\rho_{34}$ are proportional to the identity matrix. This
means that $\ket{u}$ has maximal entanglement between subsystems $(1,2)$ and $(3,4)$. Similarly, dual unitarity means
that $\rho_{13}$ and $\rho_{24}$ are also proportional to the identity, which means
maximal entanglement between subsystems $(1,3)$ and $(2,4)$. This also means that as an operator $U$ has maximal
operator entanglement between the two sites on which it acts. 

Then the only remaining bipartite entanglement measure is the
``diagonal entanglement'' of the gate, which is the entanglement between subsystems $(1,4)$ and $(2,3)$, to be computed from
$\rho_{14}$ or $\rho_{23}$. If this entanglement is also maximal, then the state $\ket{u}$ is said to be absolutely
maximally entangled, or equivalently, a perfect tensor. It can be shown that for dual
unitary gates the so-called ``entangling power'' of the gate \cite{entangling-power} depends only on the ``diagonal
entanglement'' \cite{dual-unitary--bernoulli}, and it is maximal for perfect tensors.

We say that a DU gate is non-interacting if the diagonal entanglement is zero. It can be shown that in these cases
\begin{equation}
  U=(A\otimes B)\Pe
\end{equation}
with some $A,B\in U(N)$. The resulting time evolution of the quantum circuit is trivial: there is no interaction
between qudits propagating on the different light cones.

\section{Constructions for DU gates}

\label{sec:constr}

In this Section we discuss a number of different constructions for DU gates. Most of them already appeared in the
literature, but we also include new ideas; in each case it will become clear from the text whether it is a known
construction or a new addition. In most of the examples we treat homogeneous cases, when the two sites
on which the dual unitary acts have the same dimension. However, sometimes we also consider non-homogeneous
cases. 

Before turning to the constructions let us discuss equivalence classes of DU gates.

First of all, unitarity, dual unitarity and also the
perfect tensor property are preserved by dressing with one-site unitaries. This means that if $U$ satisfies these properties, then 
for any $A,B,C,D\in U(N)$ the matrix
\begin{equation}
\tilde U= (A\otimes B)U(C\otimes D)
\end{equation}
also satisfies them.

For every DU gate $U$ the 
reshuffled gate $U^R$ is also dual unitary, this follows from the construction.  Furthermore, for each
$U$ we can define the space-reflected and time reflected gates as
\begin{equation}
  \Pe U\Pe,\qquad \mathcal{T}U\mathcal{T}=U^t=U^{t_1t_2},
\end{equation}
Both the space
and the time reflections preserve dual unitarity.
Note that we did not include a complex conjugation in the definition of the time
reflection; this has no physical meaning at this point, it is a mere convenience for the classification. The three
discrete transformations generate the group $D_4$, which is the geometric symmetry group of the square: the
transformations can 
be seen as permutations of the legs of the tensor $U$ such that the diagonals are preserved.
In contrast, for perfect tensors an arbitrary permutation of the 4 legs is an allowed transformation.

The transformations can be used to define equivalence classes of DU gates. However, different members of each class can
lead to quantum circuits with very different physical behaviour: for example fine tuned integrable gates 
can be made completely ergodic by multiplication with one-site unitaries.

For the discussion of the various constructions it is useful to focus only on the equivalence classes. Therefore, in the
subsections below we will not discuss the addition of the one-site unitaries. 

Later in Sec. \ref{sec:dupm} we also need equivalence classes of quantum circuits, such that two circuits in the same
class are 
equivalent to each other. 
Diagonal dressings of the type
\begin{equation}
  \label{diagsim}
  \tilde U= (A\otimes B)U(B^\dagger\otimes A^\dagger), \qquad A,B\in U(N)
\end{equation}
lead to equivalent circuits, because the one-site unitaries cancel on the diagonal links in the circuit. This diagonal
dressing will be used an equivalence relation in Sec.  \ref{sec:dupm}.

In some of the constructions below we will use non-homogeneous gates,  that act on two spaces with different local
dimensions. In such cases it is always understood that the gate permutes the order of the spaces in the tensor product.
For example let us choose two natural numbers $N$ and $M$ and consider the linear operator $U$ acting on
$\complex^N\otimes \complex^M\to \complex^M\otimes \complex^N$. In such a case we say that the gate is of type $(N,M)$.
The DU condition is naturally extended to these non-homogeneous cases.

\subsection{Complete classification for $N=2$}

\label{sec:N2}

For qubits the complete characterization of DU gates was performed in \cite{dual-unitary-3}. It was found that each DU
gate is equivalent to the matrix
\begin{equation}
  U=
  \begin{pmatrix}
    e^{-iJ} & & & \\
    & & -i e^{iJ} & \\
    &-i e^{iJ} & & \\
    & & &  e^{-iJ}
  \end{pmatrix},\qquad J\in \valos.
\end{equation}
This matrix can be seen as a dressing of the SWAP gate with some phases.
Conventions are chosen such that they coincide with the formulas of \cite{dual-unitary-3}.

These gates are never perfect tensors, in accordance with the fact that there is no perfect tensor with four legs
for $N=2$ (see for example \cite{AME-bounds}). 

\subsection{Dressed SWAP gates for  $N\ge 3$}

In \cite{dual-unit-param} the following construction was published for all $N$:
\begin{equation}
  \label{dressedP}
  U=D\Pe,
\end{equation}
where $\Pe$ is the SWAP gate and $D$ is a diagonal unitary matrix of size $N^2\times N^2$. Dual unitarity is easily
established, because the 
reshuffling leads to a gate of precisely this form, with the elements of $D$ reshuffled.

These gates are never perfect tensors, but they have non-zero entangling power: the phases included in the matrix $D$
 create entanglement between qudits propagating on the different light cones \cite{dual-unit-param,dual-unitary--bernoulli}.

\subsection{Direct sums}

Various types of block diagonal constructions (or direct sums) appeared in
\cite{dual-ensembles-1,dual-unitary--bernoulli,prosen-round-a-face}.
The basic idea is to consider subspaces which are kept invariant in
the diagonal direction.

Let us consider gates of type $(N,M)$ and let us split $N$ into a sum of two-nonzero integers
$N_1,N_2$. Let $U_1$ and $U_2$ be dual unitary gates of types $(N_1,M)$ and $(N_2,M)$. Then a DU gate of type $(N,M)$
is obtained as
\begin{equation}
  U=U_1\oplus U_2.
\end{equation}
Here it is understood that the product spaces are split as
\begin{equation}
  \begin{split}
      \complex^N\otimes \complex^M&=\left(\complex^{N_1}\otimes \complex^M\right)\oplus \left(\complex^{N_2}\otimes
        \complex^M\right)\\
         \complex^M\otimes \complex^N&=\left(\complex^{M}\otimes \complex^{N_1}\right)\oplus \left(\complex^{M}\otimes
        \complex^{N_2}\right),\\
        \end{split}
\end{equation}
and $U_j$ with $j=1,2$  act as $\complex^{N_j}\otimes \complex^M\to \complex^{M}\otimes
\complex^{N_j}$. As an effect of the diagonal arrangement of the splitting the invariant subspaces are also conserved
for the reshuffled matrices, and dual unitarity of $U$ follows immediately.

The splitting of the subspaces can be performed on both spaces, and it can be repeated any desired number of times.

As an extreme case of this construction we obtain controlled unitaries.  Let us consider a gate of type $(N,N)$ and
let us split the first space into a direct sum of $N$ one dimensional spaces. Then the above construction gives
\begin{equation}
  U=\Pe \left[\sum_{j=1}^N P^j_1 U^{(j)}_2\right],
\end{equation}
where $\Pe$ is the SWAP gate, $P^j=\ket{j}\bra{j}$  are projectors to local basis states, and $U^{(j)}$ are $N$
unitaries of size $N\times N$.

\subsection{Diagonal compositions}

\label{sec:diagcomp}

The product of two DU gates is not dual unitary, but a certain diagonal composition preserves dual unitarity
\cite{biunitary-qinf1}. 

\begin{figure}[t]
  \centering
 \includegraphics{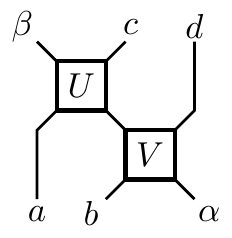}
  \caption{Diagonal composition of two non-homogeneous DU gates.}
  \label{fig:diag1}
\end{figure}

Let us choose three natural numbers $N_1,N_2$ and $M$, and consider two dual unitary gates $U$ and $V$, which are
of type $(N_1,M)$ and 
$(N_2,M)$, respectively. Then we can construct a new gate $Z$ of type $(N_1N_2,M)$ as
\begin{equation}
  \label{UV}
  Z_{ab,\alpha}^{\beta,cd}=U_{a\gamma}^{\beta c} V_{b\alpha}^{\gamma d},
\end{equation}
where the basis states of the product space $\complex^{N_1}\otimes \complex^{N_2}$ are indexed with pairs of indices
$a,b$ and $c,d$, and the Greek indices stand for the basis states in $\complex^M$.
The dual unitarity of $Z$ follows from the fact that it is a product of unitary gates in both the time
and space directions. For a pictorial interpretation see Fig. \ref{fig:diag1}.

As usually, the DU gate in eq. \eqref{UV} can be dressed by one site unitaries, which act on the local spaces of
dimension $N_1N_2$ and $M$. This allows for more freedom than just the dressing of the gates $U$ and $V$ separately. 

The steps of  construction can be repeated, and various composites can be formed. For example a large family of
(homogeneous) DU gates for $N=4$ can be constructed using the decomposition $\complex^4=\complex^2\otimes
\complex^2$. In the following we write down an explicit formula for $N=4$.

\begin{figure}[t]
  \centering
\includegraphics{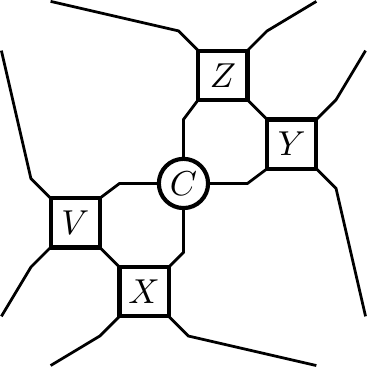}
  \caption{Diagonal composition of qubit gates, in order to obtain a DU gate for $N=4$ (representation of formula
    \eqref{U4}). The rectangles stand for  DU 
    gates, and the circle denotes an unitary operator $C$ (not necessarily DU). The resulting
    circuit is dual unitary, because it is a product of unitary operators in both the time and space directions.}
  \label{fig:diag2}
\end{figure}

We use the following notations: we regard the two-site space $\complex^4\otimes
\complex^4$ as the four-site product $\otimes_{1}^4 \complex^2$, we work with operators acting on pairs of qubits, and
when writing down the formula we omit the indices of the matrices and just denote the indices of the spaces on which
the two-qubit operators act. Let $C$ be an arbitrary two-site unitary, and $X,Y,V,Z$ be DU gates, all
corresponding to $N=2$. Then a  DU gate for $N=4$ is constructed as
\begin{equation}
  \label{U4}
  U=   Z_{23}   Y_{34}  C_{23}   V_{12}X_{23}.
\end{equation}
This operator is clearly unitary, because it is a product of unitary operators. The dual unitarity can be established
step by step, noting that the reshuffled operator (which acts in the space direction) is also a product of unitary
operators with a well defined order of their action. For a graphical interpretation of this operator see
Fig. \ref{fig:diag2}. The dual unitarity is easily read off this figure.

For $N=2$ the most general unitary and dual unitary matrices are
known explicitly (for the DU case see subsection \ref{sec:N2} above), therefore eq. \eqref{U4} gives an explicit
parametrization of a large class of DU gates for $N=4$.

These diagonal compositions can be seen as special cases of the constructions treated in the papers
\cite{biunitary-qinf1,biunitary-qinf2}, 
which deal with a more general notion of bi-unitarity. The specific formulas we gave appear to be new.

\subsection{Permutation circuits -- generalities}

\label{sec:dupm1}

A permutation matrix is a unitary operator $U$ whose matrix representation is deterministic in a chosen basis. This means
that for any 
pair of local basis states $\ket{a}\otimes \ket{b}$ we have
\begin{equation}
  U(\ket{a}\otimes \ket{b})=\ket{c}\otimes \ket{d},
\end{equation}
where $c=c(a,b)$ and $d=d(a,b)$. With some loose notation we will also write $U(a,b)=(c,d)$, in which case $U$ is
understood as a map $X^2\to X^2$ where $X=\{1,\dots,N\}$. Unitarity implies that $U$ is invertible as a map: it
describes a bijection between pairs of variables $(a,b)$ and $(c,d)$.

A quantum circuit constructed from a permutation matrix could be 
understood as a classical cellular automaton, but it is useful to regard them as special cases of a
QCA. Even though such gates are deterministic, generally they create quantum entanglement in the system
\cite{permutations-entangl}. 

The permutation map $U$ is dual unitary, if it also describes a bijection between the pairs of variables
$(a,c)$ and $(b,d)$.
Lacking a better name, such gates can be be called dual unitary permutation matrices. 
With
regards to quantum circuits they have been studied recently in \cite{dual-unitary--bernoulli}, but they appeared much
earlier in the mathematical literature. The first work discussing such maps appears to be
\cite{biunitary-permutations}, which gives a full classification up to size $N=3$.

Dual unitary permutation matrices are closely related to known objects from combinatorial designs theory. They can be
pictured as certain ``broken'' 
or ``imperfect'' orthogonal arrays or Graeco-Latin squares. The connections are as follows.

Let us list the $N^2$ quartets $(a,b,c,d)$, where it is understood that for each pair $(a,b)$ we have $(c,d)=U(a,b)$. 
Dual unitarity means that each quartet can be identified uniquely by specifying either of the following pairs of
variables: $(a,b)$, $(c,d)$, $(a,c)$, $(b,d)$. An alternative understanding is found if we compile a table of size
$N^2\times 4$ out of the quartets, such that each row corresponds to a quartet $(a,b,c,d)$. In this table we find that
there are certain pairs of columns, such that in the pair of columns every pair of numbers is present exactly once. This
is consistent with the table having $N^2$ rows. The pairs for which this
property holds correspond the pairs of variables $(a,b)$,  $(c,d)$, $(a,c)$ and
$(b,d)$. Such a table could be regarded as a ``broken orthogonal array''. The actual orthogonal arrays are similar
tables where the uniqueness property would hold for every pair of columns, or more generally, for every subset with a
given size $k$ (called the strength of the orthogonal array) \cite{OA-book}. An example for such a broken orthogonal array is shown in Table
\ref{tab:N3ex} for $N=3$. 

The dual unitary permutation matrices can be visualized also in a different way. Let us arrange the values of
the functions $c(a,b)$ and $d(a,b)$ in two matrices $C$ and $D$ of size $N\times N$, where the row index is given by $a$
and the 
column index by $b$. Such pairs of matrices are generalizations of orthogonal pairs of Latin squares, also called
Graeco-Latin squares. A Latin square is a matrix of size $N\times N$ filed with numbers $1\dots N$, such that in each
row and each column every number is present exactly one time. A pair of Latin squares $C$ and $D$ is called orthogonal,
if the pairs of numbers $(C_{jk},D_{jk})$ are all distinct. 
A permutation map $U$ is dual unitary, if the corresponding matrices $C$ and $D$ are orthogonal in the sense we just
described, and  
$C$ is only row-Latin and $D$ is column-Latin.

An example for a dual unitary permutation matrix is shown in Table \ref{tab:N3ex}. Later
in this work we will use a slightly different, more compact notation: we will combine the matrices $C$ and $D$ and for
better readability we write the pairs of numbers 
$(C_{jk},D_{jk})$ without parenthesis or separation marks.

\begin{table}
  \begin{minipage}[c]{0.2\textwidth}
    \label{aa}
    \begin{tabular}{|c|c||c|c|}
      \hline
      $a$ & $b$ & $c$ & $d$\\
      \hline
      \hline
1 & 1 &   3 & 3 \\ \hline 
1 & 2 &   2 & 3 \\ \hline 
1 & 3 &   1 & 3 \\ \hline 
2 & 1 &   3 & 1 \\ \hline 
2 & 2 &   1 & 2 \\ \hline 
2 & 3 &   2 & 1 \\ \hline 
3 & 1 &   3 & 2 \\ \hline 
3 & 2 &   1 & 1 \\ \hline 
3 & 3 &   2 & 2 \\ \hline 
  \end{tabular}
  \end{minipage}
  \begin{minipage}[c]{0.25\textwidth}
    \begin{equation*}
      C=
      \begin{bmatrix}
         3 &  2 &  1 \\ 
  3 &  1 &  2 \\ 
  3 &  1 &  2 \\ 
      \end{bmatrix}
    \end{equation*}
    \begin{equation*}
      D=
      \begin{bmatrix}
 3 &  3 &  3 \\ 
  1 &  2 &  1 \\ 
  2 &  1 &  2 \\
        \end{bmatrix}
      \end{equation*}
         \begin{equation*}
      CD=
      \begin{bmatrix}
         33 &  23 &  13 \\ 
  31 &  12 &  21 \\ 
  32 &  11 &  22 \\ 
      \end{bmatrix}
    \end{equation*}
  \end{minipage}
  \caption{Two representations of a dual unitary permutation matrix with $N=3$. On the left: representation as a broken
    orthogonal array. On the right:
representation as a pair of orthogonal squares, $C$ being row-Latin and $D$ being column-Latin. As a pair, they are
denoted together as $CD$.} 
  \label{tab:N3ex}
  
\end{table}

If a dual unitary permutation matrix is also a perfect tensor, then it corresponds to an actual orthogonal array or equivalently to a
pair of orthogonal latin squares. In the theory 
of combinatorial designs the mutually orthogonal Latin squares (MOLS) are of great interest. A MOLS is a set of Latin
squares such that any two of them are orthogonal. There is no complete classification of MOLS, and a recent work treating their
enumeration is \cite{mols-enumeration}.

If the map $U(a,b)$ is space reflection invariant, then it means that $D=C^t$ where $^t$ denotes the transpose. In the
case of the perfect maps such cases are known as self-orthogonal Latin squares, and it is known that they exist for all
$N$ except for 2, 3 and 6  \cite{self-orth-latin}. 
Space reflection invariant DU maps (which are not necessarily perfect) correspond to self-orthogonal row-Latin squares. 

For each permutation map we can also construct families of quantum gates by adding phases
\cite{dual-unitary--bernoulli}, such that 
\begin{equation}
  U(\ket{a}\otimes \ket{b})=e^{i\varphi_{cd}}(\ket{c}\otimes \ket{d}),
\end{equation}
where $c=c(a,b)$ and $d=d(a,b)$ are the classical variables and $\varphi_{cd}$ stands for a collection of $N^2$
phases. This is then a generalization of the dressed SWAP gate introduced in eq. \eqref{dressedP}.

\subsection{Finite rings}

It is well known that orthogonal latin squares can be constructed as linear maps from finite fields; this leads to known
constructions for 
perfect tensors, see for example \cite{chaos-qchann}. Now we generalize this idea to finite rings, in order to construct
dual unitary permutation matrix that are not perfect. 

Let us consider a finite commutative ring $\mathcal{R}$ of size $N$ and the linear map
\begin{equation}
  \label{ringlin}
  U(a,b)=(\alpha a+\beta b,\gamma a+\delta b),
\end{equation}
where $\alpha,\beta,\gamma,\delta\in\mathcal{R}$ are fixed parameters of the map. Even though we view $U$ as a
permutation map, it can be represented by a two by two matrix
\begin{equation}
  \begin{pmatrix}
    \alpha & \beta \\
    \gamma & \delta
  \end{pmatrix}
\end{equation}
The above permutation map is invertible if the determinant
$\alpha \delta-\beta\gamma$ is an invertible element of $\mathcal{R}$. Furthermore the map is DU if $\beta$ and $\gamma$
are also invertible elements. The map is perfect if $\alpha$ and $\delta$ are also invertible. We are looking for maps
which are DU but not perfect.

For example we consider the ring $\egesz_N$ where $N$ is not a prime. Multiplication with a non-zero element
$x\in\egesz_N$ is invertible, if $x$ is co-prime to $N$. DU maps are then given by linear maps \eqref{ringlin} where the
three elements $\beta$, $\gamma$ and $\alpha \delta-\beta\gamma$ are all co-prime to $N$.

Explicit examples are found easily. For example consider the ring $\egesz_N$ for $N=2k$ such
that $k$ is not a multiple of 3; the smallest non-trivial example is $N=4$. For such $N$ consider the space reflection
invariant map
\begin{equation}
  U(a,b)=(2a+b,a+2b).
\end{equation}
It follows from the considerations above that this map is DU, but it is not perfect.  In matrix notation it is written
as
\begin{equation}
  \begin{bmatrix}
33 & 41 & 13 & 21 \\ 
14 & 22 & 34 & 42  \\ 
31 & 43 & 11 & 23  \\ 
12 & 24 & 32 & 44  \\ 
  \end{bmatrix},
\end{equation}
where we identified $0\equiv 4$.

Perfect tensors are found most easily when the ring is actually a finite field, and
$\alpha,\beta,\gamma,\delta$ and $\alpha \delta-\beta\gamma$ are all non-zero. The simplest example for a perfect map
is found when the field is $\egesz_N$ with some prime $N\ge 3$, and the map is
\begin{equation}
  \label{lin3}
  U(a,b)=(a+b,a-b).
\end{equation}
For $N=3$ this map appeared in the works \cite{ads-code-1,chaos-qchann,dual-unitary--bernoulli}.

\subsection{Yang-Baxter maps}

A special class of dual unitary permutation matrices are those which satisfy the set-theoretical Yang-Baxter relation, which reads
\begin{equation}
  \label{YB}
  U_{12}U_{23}U_{12}=U_{23}U_{12}U_{23}.
\end{equation}
The relation is understood for operations acting on the triple product $X^3$, where $X=\{1,2,\dots,N\}$.
Introduced originally by Drinfeld \cite{Drinfeld-YB-maps} such maps were
studied in detail in the seminal paper \cite{set-th-YB-solutions}, and by now the study of Yang-Baxter maps and the
associated algebraic structures is a separate topic in mathematics (see for example the influential paper
\cite{rump-YB}). The dual unitarity condition was called 
``non-degeneracy'' in  \cite{set-th-YB-solutions}, and the class of non-degenerate solutions  of \eqref{YB} is the most
studied one. 

Cellular automata built from Yang-Baxter maps were
studied recently in \cite{sajat-superintegrable}. It was shown here that these circuits are super-integrable: they
possess an exponentially large set of local conserved quantities.

There is no general construction for dual unitary Yang-Baxter maps, but  solutions for small $N$  were enumerated in
\cite{setthYB-list}, and they are available online through the associated \texttt{github} databases (for sizes $N\le 9$ see
\cite{YB-gap}).

\subsection{Graph states}

So-called graph states were used in \cite{four-AME,AME-graph,AME-graph-2} to construct perfect tensors. In a concrete
representation we obtain tensors, such that all components have equal magnitude but varying phases, in order to produce
maximal entanglement. These ideas can be applied naturally also to DU gates, via the operator-state
correspondence. We obtain special complex Hadamard matrices \cite{complex-hadamard}, which appear as generalizations of
the discrete Fourier transform. A number of examples already appeared in the literature. The DU gate of the kicked Ising
model is in this class 
\cite{dual-unitary-1,dual-unitary-2,dual-kicked,sarang-lamacraft-kicked-DU}. Recently generalizations also
appeared in \cite{dual-unitary--bernoulli} which were called ``quantum cat maps''. Here we give one slight
generalization of the class given in \cite{dual-unitary--bernoulli}. We will see that these gates are also related to
the permutation maps. 

Let us assume that
\begin{equation}
  U^{ij}_{kl}=\frac{1}{N} \omega^{F(i,j,k,l)},
\end{equation}
where $\omega$ is the first root of unity given by $\omega=e^{2\pi i/N}$ and $F(i,j,k,l)\in \egesz$. It is natural to
interpret the variables $i,j,k,l$ and the values of $F$ to be elements of the ring $\egesz_N$. We do not assume $N$ to
be prime, so $\egesz_N$ is not necessarily a field.

We are looking for $F$ functions which are bi-linear:
\begin{equation}
  \label{bili}
  F(i,j,k,l)=\alpha_1ij+\alpha_2 kl+\beta_1ik+\beta_2 jl+\gamma_1 il+\gamma_2 jk.
\end{equation}
Linear pieces would correspond to the action of one-site unitaries, therefore we do not add them in this Ansatz. The
coefficients $\alpha_{1,2}$, $\beta_{1,2}$ and $\gamma_{1,2}$ are also assumed to be elements of $\egesz_N$.
A bi-linear function of this form can be represented by a graph with 4 vertices, and 6 weighted edges between them, such
that the weights are given by the coefficients above. Hence the name ``graph states''.

The unitarity and dual unitarity conditions are
\begin{equation}
  \frac{1}{N^2}\sum_{kl} \omega^{F(i,j,k,l)-F(m,n,k,l)}= \delta_{im}\delta_{jn}
\end{equation}
and
\begin{equation}
   \frac{1}{N^2}\sum_{kl} \omega^{F(i,j,k,l)-F(i,m,k,n)}=
   \delta_{jm}\delta_{ln}.
\end{equation}
It is our goal to fix the six numerical coefficients $\alpha_{1,2},\beta_{1,2},\gamma_{1,2}$ such that both relations
are satisfied. 
For any $x\in \field_N$ we have the obvious identity
\begin{equation}
  \sum_{k=1}^N \omega^{kx}=N\delta_{x,0}.
\end{equation}
Our goal is to ensure that in the summations above the non-zero terms always correspond to the desired
identities.

The unitarity condition holds if the equations
\begin{equation}
  \beta_1 (i-m)+\gamma_2 (j-n)=0,\qquad \gamma_1(i-m)+\beta_2(j-n)=0
\end{equation}
imply $i=m$, $j=n$. The dual unitarity holds if the equations
\begin{equation}
  \alpha_1(j-m)+\gamma_1(l-n)=0,\qquad \gamma_2(j-m)+\alpha_2(l-n)=0
\end{equation}
imply $j=m$, $l=n$. This is satisfied in the finite ring $\egesz_N$ if the $2\times 2$ matrices corresponding to the
equations above are
invertible, which means that $\beta_1\beta_2-\gamma_1\gamma_2$ and $\alpha_1\alpha_2-\gamma_1\gamma_2$ have to be
invertible elements of $\egesz_N$ (co-prime to $N$).

Examples are found already for $N=2$: the self-dual kicked Ising model \cite{dual-unitary-1} corresponds to  
\begin{equation}
  \label{kick}
  \alpha_1=\alpha_2=\beta_1=\beta_2=1,\text{ and }\gamma_{1,2}=0.
\end{equation}
This choice of parameters gives a DU
gate for every $N$. The corresponding graph state is known in quantum information theory as the cluster state or ``C
state'' \cite{four-AME}.

Generalizations of this case  called ``quantum cap maps'' were published in
\cite{dual-unitary--bernoulli}; these solutions also have two parameters equal to zero as in the kicked Ising case. A
representative can be chosen as
\begin{equation}
  \label{P}
  \gamma_1=\gamma_2=0,\qquad \alpha_1=\alpha_2=\beta_1=1,\qquad \beta_2=-1.
\end{equation}
This is known as the ``P state''  in quantum information theory \cite{four-AME}.

The graph state construction gives a perfect tensor if $\alpha_1\alpha_2- \beta_1\beta_2$ is also invertible.  It is
known that the P state is perfect for every odd $N$, this follows simply from our requirements.  On the other hand, the
C state is never perfect.

Let us now consider a special case when $\alpha_1=\alpha_2=0$. In this case the gate is equivalent to a linear map
constructed from the finite field $\egesz_N$. To see this we use the one-site unitary $F$, which is the matrix of the
discrete Fourier transform given by
\begin{equation}
  F_{jk}=\frac{1}{\sqrt{N}}\omega^{-jk}.
\end{equation}
Computing
\begin{equation}
\tilde U=  (F\otimes F)U
\end{equation}
we see that $\tilde U$ is a permutation map of the form \eqref{ringlin} with a $2\times 2$ matrix given by
\begin{equation}
  \begin{pmatrix}
    \beta_1 & \gamma_1 \\
    \gamma_2 & \beta_2
  \end{pmatrix}.
\end{equation}

\section{Ergodicity of DU permutation circuits}

\label{sec:dupm}

Now we consider the ergodicity properties of the circuits built from DU permutation maps. In this Section we focus on DU gates
which are not perfect maps; perfect permutation maps will be considered separately in the next Section.

We analyze the in-equivalent  circuits for small local dimensions $N=2$ and $N=3$. We consider two physical properties of
these systems: the number of local conserved charges as a function of the range of the charge density, and the orbit
lengths in the 
corresponding classical cellular automaton. Before turning to the concrete examples let us first discuss these two properties.

It was already in explained in Section \ref{sec:corr} that in DU circuits all conserved charges are gliders: the time
evolution of the charge densities is a mere translation to the left or to the right with constant speed, without any
operator spreading. 
In most previous works the ergodicity of the DU circuits was characterized based on the one-site
density matrices: it was implicitly stated that a DU circuit is non-ergodic iff there is at least one one-site
glider. At the same time, it was mentioned in \cite{dual-unitary-3} that there can be models where the one-site density
matrices are fully chaotic, but gliders can appear within the Hilbert spaces of the multi-site operators (see also the
so-called ``still gliders'' wilt multi-site support found in \cite{thermalizing-boundaries}). In this work we show
concrete examples for this behaviour.   

It was also mentioned in Section \ref{sec:du}, that if a DU 
circuit has at least one conserved charge, then it has infinitely many.
In such a case a
relevant question is, whether there are additional gliders at longer range which are algebraically independent from
products of shorter gliders. As far as we know this question has not yet been investigated in the integrable DU
circuits. Partial results are found in the work \cite{sajat-superintegrable}, which treated DU circuits from Yang-Baxter
maps. In these models an exponentially large set of 
gliders was constructed, such that the multi site gliders are indeed operator products of the three-site gliders (this
was not explicitly stated in \cite{sajat-superintegrable} but it is clear from the construction). However, even in these
cases it is not clear whether there are always one-site gliders, or additional charges on top of those following from
the Yang-Baxter structure.

Below we collect some numerical data about the number of gliders for various models and operator lengths. The number of
gliders with a maximal range $2\alpha-1$ can be found most easily by looking at the eigenvalues of 
the transfer matrix $t_\alpha$ written earlier in \eqref{talpha}. We numerically diagonalize the $t_\alpha$ for various
permutation circuits and ranges $\alpha\le 4$. Each time we publish the number of eigenvalues which are unity or roots
of unity, subtracting the trivial case corresponding to the identity operator. The algebraic independence of
these operators is a separate question, which should be investigated on a case by case basis; below we just publish the
overall numbers and the conclusions that can be drawn from this.

Besides the existence of gliders the other physical property we look at is the average length of the orbits in the dual
unitary permutation matrices in a finite
volume $L$ (for similar earlier studies see for example \cite{sarang-rule150,sarang-rule54,sajat-superintegrable}). If $\ket{\Psi}$ is 
a product state in the 
local basis, then the set of product states $\ket{\Psi(t)}, t=0,1,2\dots$ is called the orbit of $\ket{\Psi}$. The time
evolution is invertible, and the configuration space is finite for finite $L$, therefore all orbits are periodic, and we
say $n>0$ is the orbit length of the product state 
$\ket{\Psi}$ if it is the smallest number $n$ such that $\ket{\Psi(2n)}=\ket{\Psi(0)}$. We included a factor of 2 in
this definition, because the Floquet cycle has period 2, and orbits can be defined only for completions of a Floquet cycle.  We
define $\bar n$ as the average orbit length computed over all states of the system. 

In the numerical computations the observed average orbit lengths are only approximations of the true value, because for
larger volumes it is not possible to cover the complete configuration space. Therefore our numerical method was the
following: we computed the orbit lengths for 200 randomly chosen initial configurations, and computed the average. In
order to test the accuracy of these results, we run multiple ($\simeq 10$) instances of these averaging procedures and
computed the variance of the results obtained. Whenever we performed these checks we found a variance of less than 3\%
in the average on logarithmic scale.

The orbit lengths carry information about the global ergodicity properties of these deterministic circuits. In the most
ergodic cases the average orbit lengths grow together with the dimension of the Hilbert space. In such a case
 $\bar n\sim N^L$: this means that there are a few giant orbits (or possibly just a single giant orbit) covering a
 finite fraction of the configuration space, with the rest covered by smaller orbits. 
On the other hand, conserved charges foliate the configuration space, and this leads to shorter orbit lengths. Therefore
we expect that the number 
of gliders should be reflected in the average orbit lengths.

During the enumeration of  the models we consider two circuits to be equivalent if their two-site gates are related to
each other by combinations of time or space reflection, 
and the diagonal similarity transformations \eqref{diagsim}.

\subsection{$N=2$}

For $N=2$ there are 5 non-equivalent DU permutation circuits. 3 of them are non-interacting: they are $U=\Pe$,
$U=\sigma^x_1\Pe$ 
and $U=\sigma^x_1\sigma^x_2\Pe$. The remaining two cases are
\begin{equation}
U_1=\begin{bmatrix}
 12 &  21 \\ 
  11 &  22 \\ 
\end{bmatrix}\quad\text{and}\quad
U_2=\begin{bmatrix}
 21 &  12 \\ 
  22 &  11 \\ 
 \end{bmatrix}.
\end{equation}
Here we used the matrix notation explained in Section \ref{sec:dupm1}.

These gates can be understood as combinations of the CNOT and SWAP gates, and also a single site spin flip.
Both models have one left moving and right moving glider already for range $\alpha=1$, so they are not
ergodic. A direct check of eq. \eqref{YB} shows that $U_1$ and $U_2$ are not Yang-Baxter maps, nevertheless they appear
integrable: the orbit lengths are found to be at most linear in $L$ in both models. 

\subsection{$N=3$}

Dual unitary permutation matrices with $N=3$ offer more possibilities and some surprises. For $N=3$ we performed a brute
force computer search and found 227 non-equivalent models. Each one of the 227 models can be  
connected to one of the 18 bi-unitary permutations found in \cite{biunitary-permutations}, by adding a SWAP gate and
further one-site permutations. However, the single permutations can drastically change the physical properties of the
circuits. These permutation matrices were already investigated in the paper \cite{dual-unitary--bernoulli}, which considered the
various possibilities for the entangling power and the behaviour for the one-site operators. Our contribution is that 
we consider the higher charges and the orbit lengths of the resulting circuits. 

We computed the numbers of gliders with ranges $\alpha=1,2,3,4$ for all 227 models, and for a selected set of models we
also considered the orbit lengths. For most models we observed the expected behaviour, that the information
gained from the one-site operators completely determines the physical behaviour. However, for a smaller number of models
we also found some surprises: Sometimes the models do not have any gliders for $\alpha=1$ or even $\alpha=2$, but
gliders start to appear at longer ranges. As far as we know this has not yet been observed in the literature.

Regarding orbit lengths we observed the expected behaviour: as a rule of thumb, the more charges a model has, the shorter its
classical orbits are. 

Publishing the data corresponding to all 227 models seems pointless, instead we focus on concrete examples. The numbers
of gliders are summarized in Table \ref{tab:gl}, and the orbit lengths are discussed in the text.

\begin{table}[t]
  \centering
  \begin{tabular}{|c||c|c|c|c|}
    \hline
$\alpha$    & 1 & 2 & 3 & 4  \\
    \hline
    \hline
  $C_1$, $C_2$  & 0  &  0 & 0  & 0  \\
    \hline
$I_1$      & 4 & 40 & 364 & 3280 \\
    \hline
$I_2$      & 8 & 80 & 728 & 6560 \\
    \hline
$E_1$    & 0 & 2   & 11  & 54 \\
      \hline
 $E_2$     & 0 & 0 & 1 & 3 \\
    \hline 
  \end{tabular}
  \caption{Total number of non-trivial gliders for a given range, for 6 selected models at $N=3$. The table shows the number of
    eigenvalues $|\lambda|=1$ of the transfer matrices $t_\alpha$ plus the same number for the space reflected version,
    minus the two trivial eigenvalues corresponding to the identity matrices. We did not check  the
    algebraic independence of the gliders. }
  \label{tab:gl}
\end{table}

\begin{figure}[ht]
  \centering
  \begin{minipage}{0.49\linewidth}
    \includegraphics[scale=0.18]{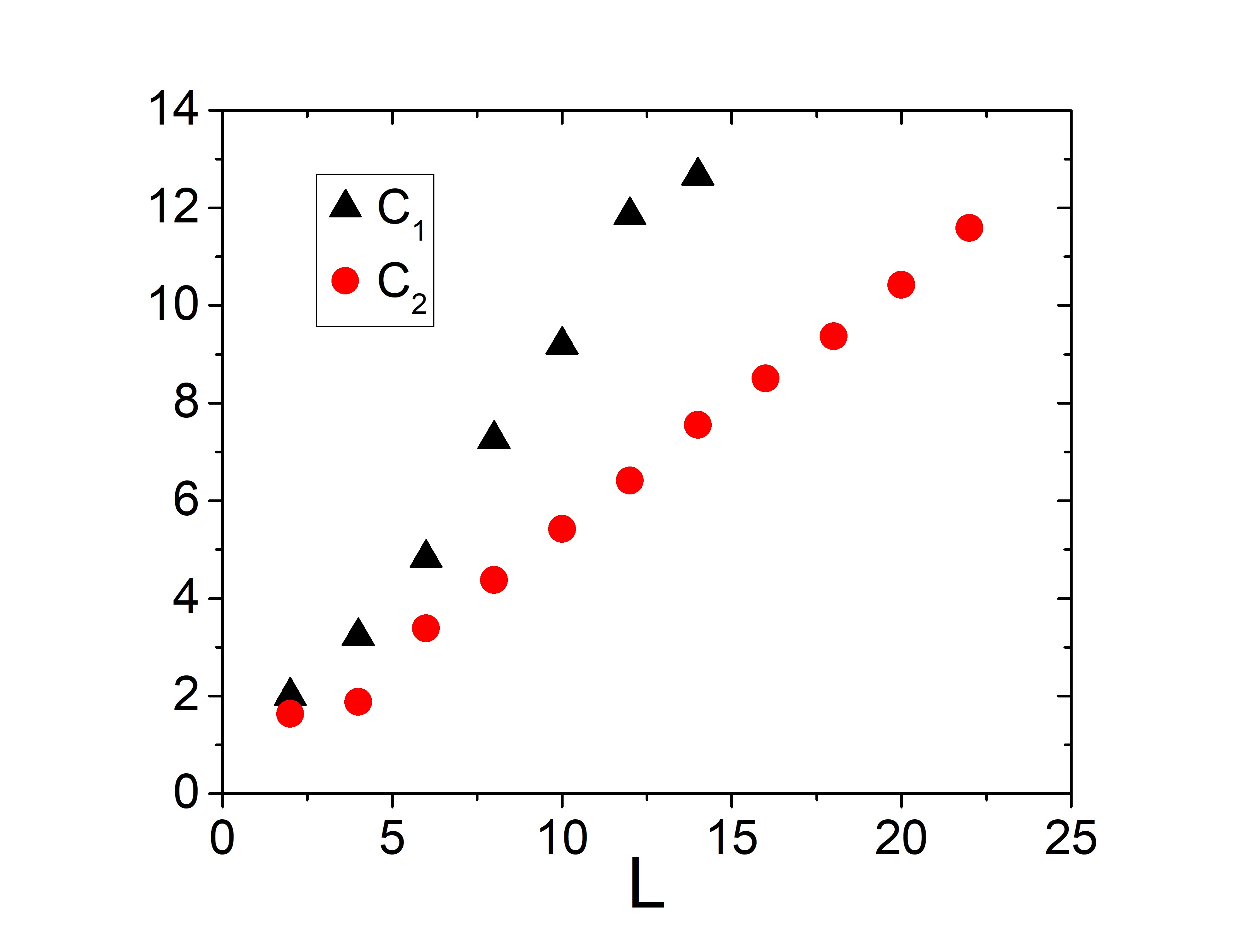}
 \end{minipage}
  \begin{minipage}{0.49\linewidth}
     \includegraphics[scale=0.18]{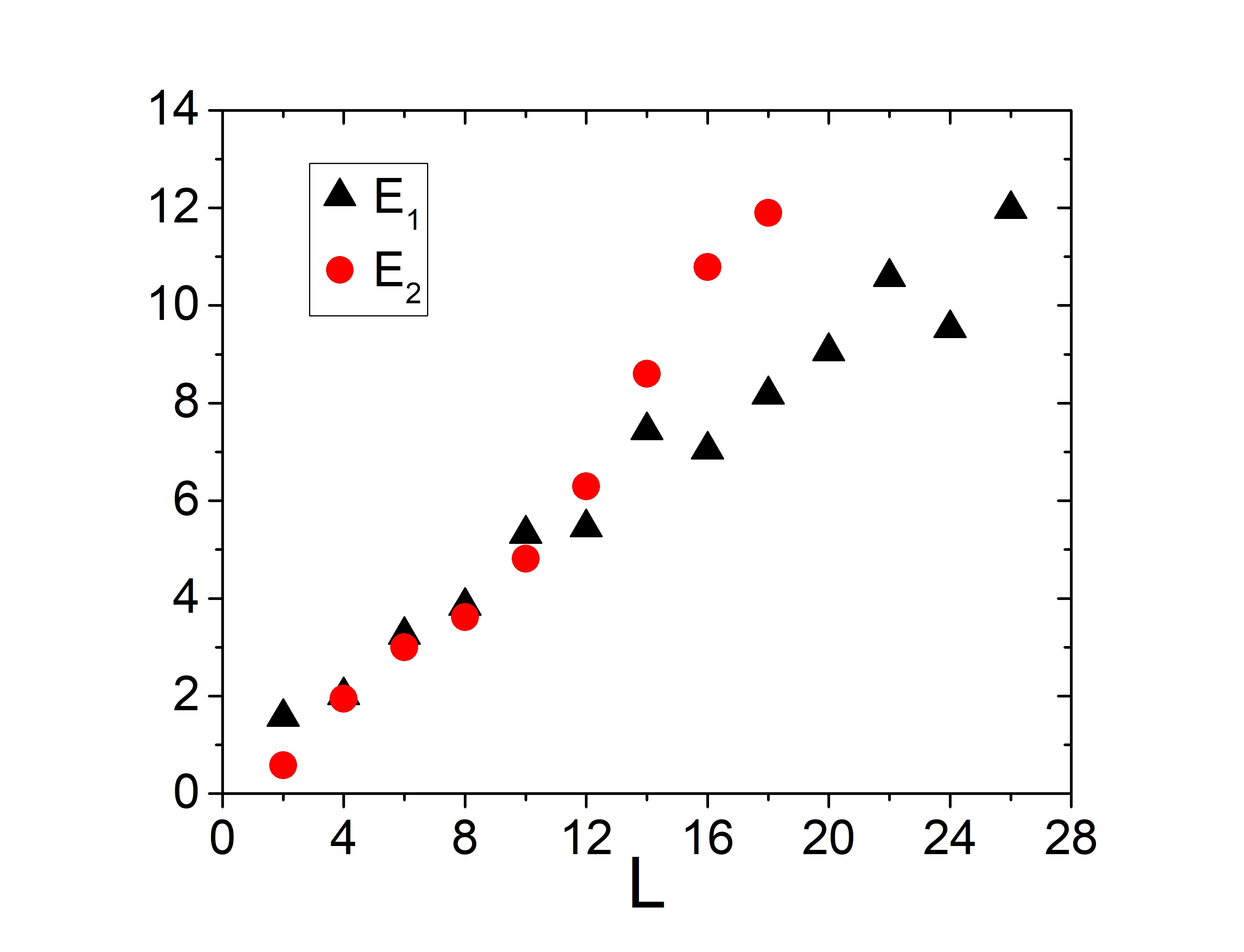}
  \end{minipage}
  \caption{Average orbit lengths  for four selected models; we plotted $\log_3 \bar n$ as a function of the volume. The
    labels of the models 
    correspond to the matrices given in the main text.}
\label{fig:orbit0}
\end{figure}

As first ergodic examples we consider the models given by 
\begin{equation}
C_1=\begin{bmatrix}
 32 &  22 &  13 \\ 
  23 &  33 &  12 \\ 
  21 &  31 &  11 \\ 
\end{bmatrix},
 \quad\text{and}\quad
C_2=\begin{bmatrix}
 13 &  23 &  33 \\ 
  31 &  12 &  22 \\ 
  32 &  21 &  11 \\ 
 \end{bmatrix}.
\end{equation}
Neither of these two models has any glider up to range $\alpha=4$. The orbit lengths grow as
$\bar n\approx 3^L$ and $\bar n\approx 3^{L/2}$, for $C_1$ and $C_2$, respectively; they are plotted in
Fig. \ref{fig:orbit0} on a logarithmic scale. 

The model $C_1$ is as close to perfectly
ergodic as possible: for each $L$ we find one giant orbit, covering a finite fraction of the configuration space, and
a number of shorter orbits. 
For $C_2$ we find that the average
orbit lengths scale with the square root of the size of the Hilbert space: this decrease could be a result of a hidden
discrete symmetry, but we did not find an explanation yet.

As examples for integrable cases we present the following two maps:
\begin{equation}
  I_1=
  \begin{bmatrix}
 11 &  31 &  21 \\ 
  12 &  23 &  33 \\ 
  13 &  22 &  32 \\ 
 \end{bmatrix}
 \quad\text{and}\quad
  I_2=\begin{bmatrix}
 11 &  21 &  31 \\ 
  12 &  32 &  22 \\ 
  13 &  23 &  33 \\ 
 \end{bmatrix}.
\end{equation}
Both maps are interacting: they have non-zero entangling power. 
For the average orbit lengths we find $\bar n\approx L$; also the maximal orbit length is found to increase only linearly
with the volume. Direct computation shows that $I_1$ is a Yang-Baxter map, this explains its integrability
\cite{sajat-superintegrable}. $I_2$ is not a Yang-Baxter map, but it has a very simple structure: it is a controlled
unitary, which we write as 
\begin{equation}
I_2=\Pe\left[  P^{(1)}_1M_2+P^{(2)}_1+P^{(3)}_1   \right],\quad M=
\begin{pmatrix}
  0 & 0 & 1\\
  0 & 1 & 0\\
  1 & 0 & 0
\end{pmatrix}.
\end{equation}
Here the one-site projectors are as before $P^{(j)}=\ket{j}\bra{j}$ and $M$ is one-site permutation matrix. The maximal
number of gliders for $\alpha=1$ is $2(N^2-1)$ and it appears that $I_2$ has half of them, and $I_1$ has a quarter of them.

We also found a total number of 30 non-equivalent circuits which have no gliders at range $\alpha=1$ but a varying number of
gliders for $\alpha=2$. None of these gates is a Yang-Baxter map. In addition, we found 3 non-equivalent circuits that
have no gliders for $\alpha=1,2$ but a single glider for $\alpha=3$.
Examples for models with gliders appearing at $\alpha=2$ and $\alpha=3$ are given by
\begin{equation}
  \label{E12}
  E_1=
  \begin{bmatrix}
 32 &  22 &  13 \\ 
  31 &  21 &  12 \\ 
  23 &  33 &  11 \\ 
 \end{bmatrix}
  \quad\text{and}\quad
  E_2=
  \begin{bmatrix}
 11 &  32 &  21 \\ 
  33 &  13 &  23 \\ 
  12 &  31 &  22 \\ 
 \end{bmatrix}.
\end{equation}
For $E_1$ we find a total number of 2 gliders for $\alpha=2$ (both are right-moving),
and for $E_2$ we have a single glider with $\alpha=3$.
The growth of the orbit lengths in these models is plotted on Fig. \ref{fig:orbit0}, again in logarithmic scale. For
$E_1$ we see a clear exponential growth, with a non-integer exponent. Data on $E_2$ is less clear, and likely larger
volumes would be needed to establish the growth exponent.

The models given by $E_1$ and $E_2$
should be called ``integrable'', because after all they have an infinite set of gliders. 
The algebraic reason for their  integrability is not known at the moment. Whether or not integrability is truly
``accidental'' here, or whether there a deeper reason behind it, is a question for further research.

\section{Ergodicity properties of DU circuits with perfect maps}

\label{sec:perfect}

In this Section we investigate circuits with perfect permutation maps for $N=3,4,5$. As discussed above, such maps
correspond to orthogonal Latin squares.  As quantum gates they are perfect tensors and they have maximal entangling
power. As discussed in \cite{dual-unitary--bernoulli} they are maximally chaotic on the level of local correlation
functions. Furthermore, they are strong scramblers: the perfect property can be used to analytically prove the maximal
operator spreading \cite{chaos-qchann}. 

The work  \cite{chaos-qchann} considered the perfect map given by \eqref{lin3}, which is a linear map over finite
fields. At the first sight this linearity has nothing to do with the ergodicity of the resulting quantum circuit: the
map is a perfect tensor, and the circuit  is locally maximally chaotic. Nevertheless, it was observed in
\cite{chaos-qchann} that there are global signs of the integrability: the recurrence time of the system
shows unexpected behaviour. We discuss this phenomenon in detail.

The recurrence time $T$ of the deterministic circuit is defined as the smallest
non-zero  integer such that $\ket{\Psi(2T)}=\ket{\Psi(0)}$ for all initial states $\ket{\Psi(0)}$. $T$ is the least
common multiple of the orbit 
lengths, and it is the classical analogue of the Poincair\'e recurrence time. For ergodic systems $T$ is expected to grow doubly
exponentially \cite{chaos-qchann}. On the other hand, it 
is known that in some integrable models it grows only polynomially. For example in the case of dual unitary Yang-Baxter
maps we have $T=L/2$, and for other Yang-Baxter maps quadratic or cubic growth was  also demonstrated in
\cite{sajat-superintegrable}. 

\begin{figure}[ht]
  \centering
  \begin{minipage}{0.49\linewidth}
    \includegraphics[scale=0.18]{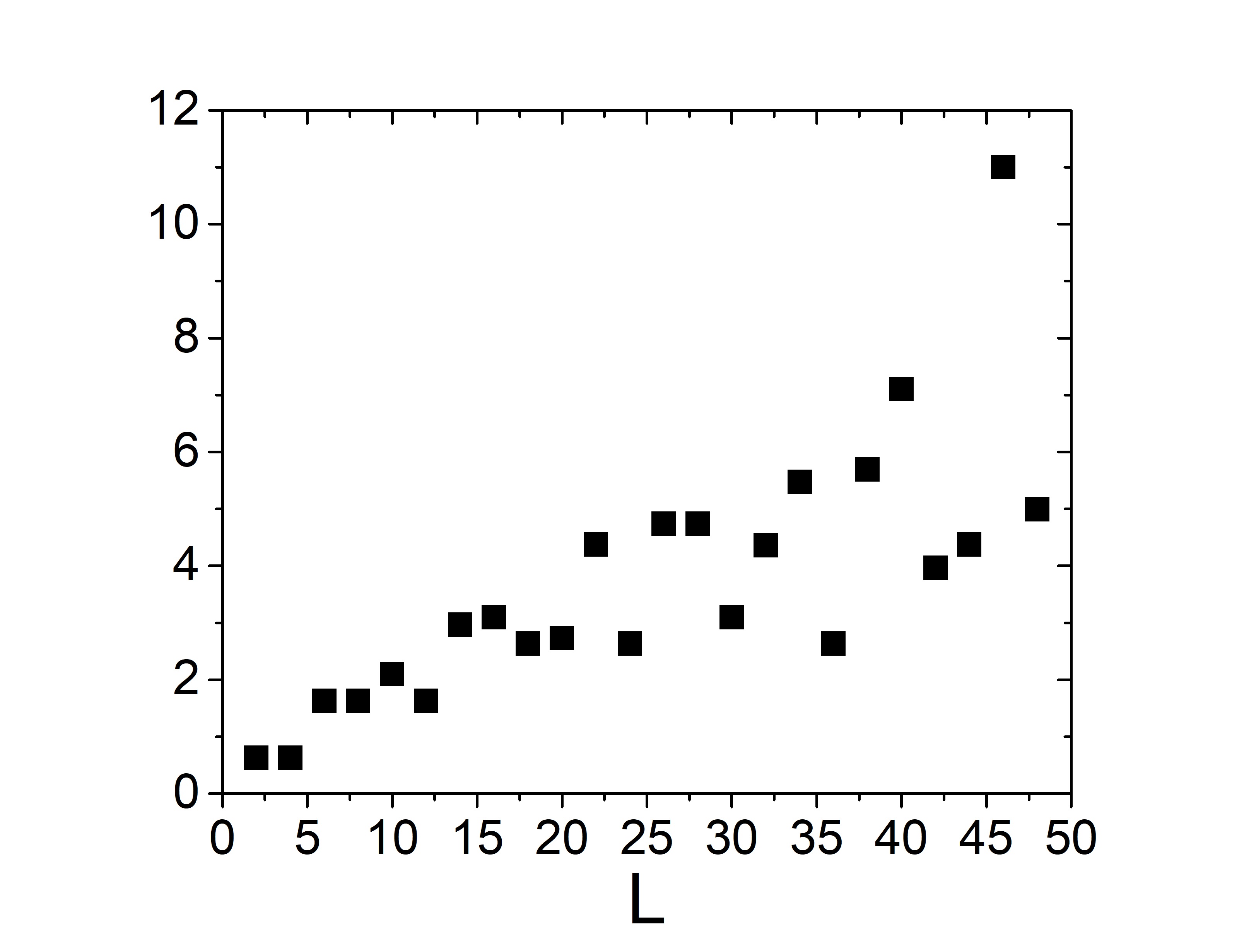}
 \end{minipage}
  \begin{minipage}{0.49\linewidth}
     \includegraphics[scale=0.18]{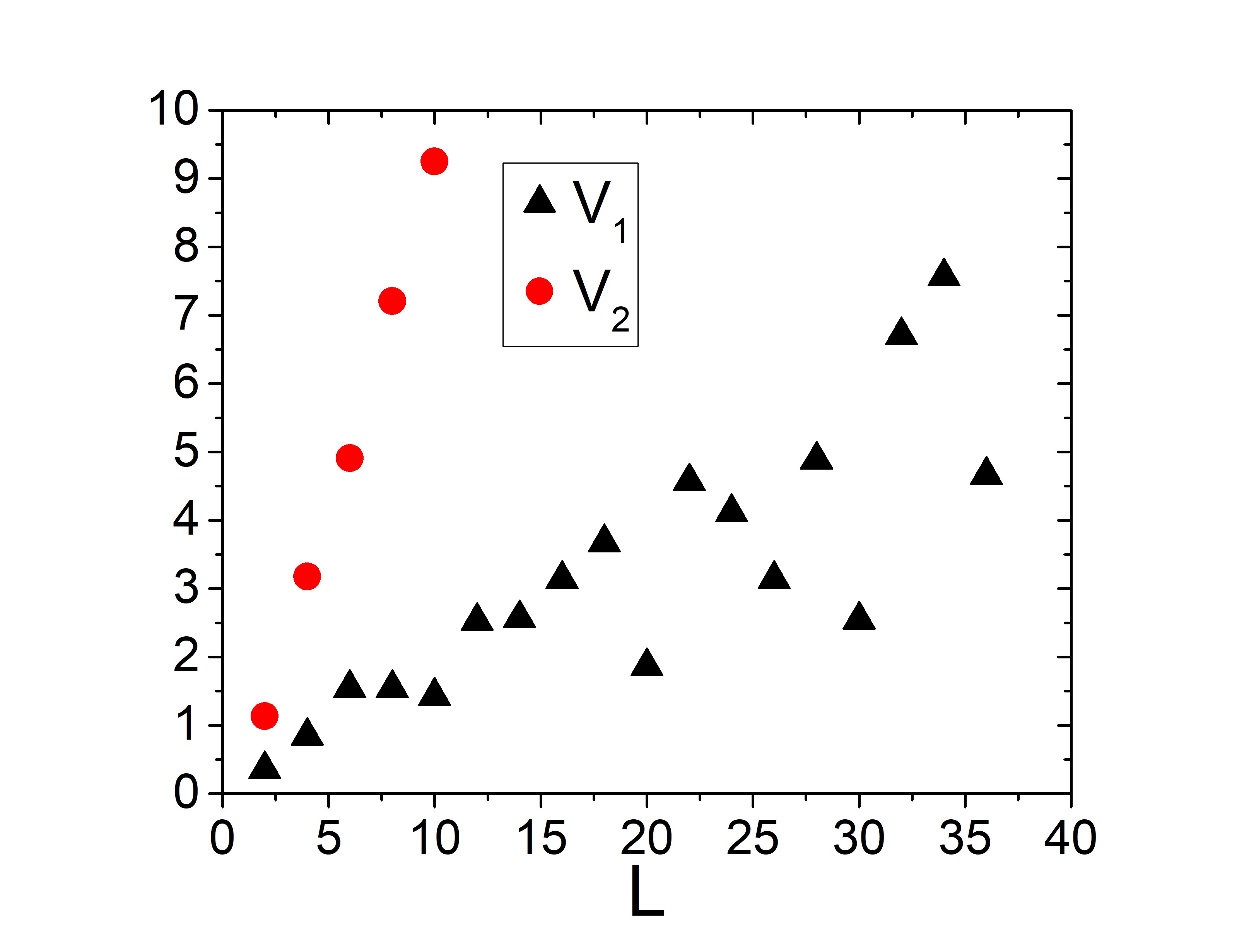}
  \end{minipage}
  \caption{On the left: Exact recurrence times of the linear model over $\field_3$; this is the plot of $\log_3(T(L))$
    with $T(L)$  presented in Table \ref{tab:field3}. On the right:  Average orbit lengths for two models at $N=5$: the linear model
    given by $V_1$ and the dressed model given by $V_2$. We plotted $\log_5(\bar n)$, and for $V_2$ we observe a linear
    function with slope $\approx 1$, thus the model is maximally ergodic.}
\label{fig:orbit0b}
\end{figure}

It was noticed in \cite{chaos-qchann} that for the perfect tensor map \eqref{lin3} over $\field_3$ the recurrence time $T$ grows at
most exponentially with the volume, however, $T$ shows a very irregular behaviour, depending very strongly on the prime
factorization of the number $L$. It was noted in \cite{chaos-qchann} that for volumes $L=2\cdot 3^m$ with some
$m=1,2,\dots$ the growth is linear: $T=2L$, but for other values $L=2q$ with some prime $q$ it grows exponentially with
an unknown exponent. Numerical data for the recurrence times is presented in Table \ref{tab:field3}, and for
completeness we also plot in Fig. \ref{fig:orbit0b}.
We can see
certain values of $L$ where $T$ appears to be growing exponentially, just to be followed by other values of $L$ with
unexpectedly small recurrence times. 
This behaviour is in stark contrast with the maximal ergodicity seen in the local quantities.

\begin{table}
  \centering
  \begin{tabular}{*{29}{|c}|}
  \hline
$L$&   2 &      4 &       6 &       8 &      10  &     12  &     14  &     16  &     18  &     20  &     22   &    24  &
                                                                                                                     26
  &     28 &      30\\
  \hline
  $T$ &  4    &    4    &   12 &      12  &     20  &     12  &     52  &     60  &     36   &    40  &    244  &     36  &    364  &    364   &    60\\
\hline
  \end{tabular}
  \medskip
  
  \begin{tabular}{*{10}{|c}|}
  \hline
$L$&   32     &  34 &      36  &     38  &     40  &     42  &     44  &     46   &    48      \\
  \hline
  $T$&  240  &    820  &     36 &    1036  &   4920  &    156  &    244 &  354292  &    180  \\
\hline
  \end{tabular}
  \caption{Recurrence times in the linear model over $\field_3$ for small values of $L$.}
  \label{tab:field3}
\end{table}

The determination of the function $T(L)$ is an interesting question on its own, which can be formulated as a problem 
in linear algebra over finite fields. More generally, let us choose a prime $N=p\ge 3$ and
consider the vector space $(\field_p)^{L}$ with some even volume $L$.
Consider the linear map
\begin{equation}
  \label{linp}
  U(a,b)=(a+b,a-b)
\end{equation}
over $(\field_p)^{2}$. The extension to a Floquet circuit gives the update rule $\VV$, which can be represented
as a linear operator acting on this vector space:
\begin{equation}
  \label{VVdef}
  \VV=AB.
\end{equation}
Here $A$ is the block diagonal matrix
\begin{equation}
  A=
  \begin{pmatrix}
    1 & 1 & &  & \\
    1 & -1 && & \\
   && \ddots & & \\
    & &&1 & 1\\
    & &&1 & -1\\
  \end{pmatrix}.
\end{equation}
and $B=CAC^{-1}$ with $C$ being the cyclic shift operator on $(\field_p)^{L}$. The matrices are of size $L\times L$ with
elements in $\field_p$. The order of the product in \eqref{VVdef} is actually different from the choice of
\eqref{VVdef0}, but this does 
not change the physical properties, and it is chosen to conform with the conventions of the Appendix.

The recurrence time $T$ is defined as the smallest non-zero integer such that
\begin{equation}
  \VV^T=1
\end{equation}
as a matrix over $(\field_p)^{L}$. 

The determination of $T(L)$ (together with an actual solution of the time evolution)  requires the use of the
Fourier transform over extensions of the finite field $\field_p$. This is treated as a separate problem in the Appendix,
written by Roland Bacher and Denis Serre. The Appendix includes a rigorous proof of the linear growth of $T(L)$ for
$L=2p^m$ and it is also proven that the maximal growth of $T(L)$ is only exponential in $L$.

Both the theoretical treatment and also the numerical data show that these linear maps behave essentially the same way
for all $p$. Furthermore, it is not at all crucial which linear map we choose, as long as it is a perfect map: there can
be some differences in the actual values of $T(L)$, but the overall characteristics are the same. Furthermore,
essentially the same behaviour is also seen if the perfect map is linear over a finite field of order $p^m$ with some $m>2$. 

The global non-ergodicity and the solvability are consequences of the linearity of the map, but strangely this is seen
only at the ``classical level'', and the local indicators of quantum chaos are signalling completely non-ergodic
behaviour. The observed phenomenon is very similar to the behaviour of the discrete version of Arnold's cat-map, whose
recurrence time was studied long time ago by Dyson and Falk \cite{cat-map}.

It was already suggested in \cite{chaos-qchann} that a truly ergodic perfect quantum circuit could be constructed if the linear
maps are dressed with one-site unitaries, for example with some rotations with non-rational parameters. We considered
this idea within the class of dual unitary permutation matrices: we added one-site permutations to the outgoing legs of
the linear map, and 
investigated the orbit lengths of the circuit thus obtained.  For $N=3$ this idea does not give anything new, because
every permutation of three elements can be described by a linear function in $\field_3$, thus dressing with one-site permutations
would not give a non-linear map. Instead, we considered the next prime $N=5$ and the maps
\begin{equation}
  \label{USS}
  \tilde U=(S_1\otimes S_2)U,
\end{equation}
where $U$ is the same linear map \eqref{linp} as before, but $S_1$ and $S_2$ are permutations that are not linear maps
in $\field_5$. We observed the expected behaviour: the addition of just one ``non-linear'' permutation is enough to
guarantee full ergodicity.

To be concrete, let us consider the map \eqref{linp} over $\field_5$, represented by the matrix 
\begin{equation}
V_1=\begin{bmatrix}
 11 &  25 &  34 &  43 &  52 \\ 
  22 &  31 &  45 &  54 &  13 \\ 
  33 &  42 &  51 &  15 &  24 \\ 
  44 &  53 &  12 &  21 &  35 \\ 
  55 &  14 &  23 &  32 &  41 \\ 
 \end{bmatrix}.
\end{equation}
Here we chose a convention so that the pair $(1,1)$ is kept invariant (vacuum configuration), so that the elements
$0\dots 4\in \field_5$ correspond to our basis states $\ket{1},\ket{2},\dots,\ket{5}$ in this order.
Let us now consider the permutation $S$ given by
$(1,2,3,4,5)\to (1,5,3,2,4)$. It is easy to check that this permutation is not a linear map over $\field_5$. Now we
perform this permutation on the first outgoing variable of $V_1$ according to \eqref{USS}. This leads to the matrix
\begin{equation}
V_2=  \begin{bmatrix}
 11 &  55 &  34 &  23 &  42 \\ 
  52 &  31 &  25 &  44 &  13 \\ 
  33 &  22 &  41 &  15 &  54 \\ 
  24 &  43 &  12 &  51 &  35 \\ 
  45 &  14 &  53 &  32 &  21 \\ 
 \end{bmatrix}.
\end{equation}
The average orbit lengths for the models given by $V_1$ and $V_2$ are plotted in Fig. \ref{fig:orbit0b}. For $V_1$ we
observe the same pattern as earlier for the linear map: the orbit lengths depend crucially on the prime factorization of
the volume, and the recurrence time grows only exponentially. For $V_2$ on the other hand we see the expected maximal
ergodicity: the average orbit lengths grow approximately as $5^L$.

For completeness we also considered perfect maps that have absolutely no relation with finite fields. It is known that the smallest
local dimension at which such a map appears is $N=7$, in which case there are 6 non-equivalent perfect maps such that
only one of them is related to $\field_7$ and the other 5 are unrelated \cite{mols-enumeration,MOLS-online}. An example
for such an unrelated map is  given by \cite{MOLS-online}
\begin{equation}U=
  \begin{bmatrix}
 11 &  22 &  33 &  44 &  55 &  66 &  77 \\ 
  24 &  16 &  41 &  35 &  67 &  73 &  52 \\ 
  36 &  45 &  12 &  23 &  71 &  57 &  64 \\ 
  47 &  51 &  65 &  72 &  26 &  34 &  13 \\ 
  53 &  37 &  76 &  61 &  14 &  42 &  25 \\ 
  62 &  74 &  27 &  56 &  43 &  15 &  31 \\ 
  75 &  63 &  54 &  17 &  32 &  21 &  46 \\ 
 \end{bmatrix}.
\end{equation}
In this case we also observed complete ergodicity: for the small volumes accessible to numerical treatment  we observed
orbit lengths comparable to $7^L$. 

\section{Discussion}

In this work we studied the dynamics of the DU circuits with permutation maps, and discussed two unexpected phenomena:
1) There are models, where conserved charges appear only for multi-site operators (concrete examples were given by the matrices $E_1$ and $E_2$ in \eqref{E12}). 2) There are models, which appear
maximally chaotic when looking at the local quantum 
mechanical behaviour, whereas they appear in fact integrable/solvable if good classical coordinates are chosen (these
are the DU gates obtained from the linear maps, see Section \ref{sec:perfect}). Both
phenomena tell us that caution needs to be exercised when classifying quantum cellular automata based on the common
indicators of quantum chaos.

It would be important to see, whether these two phenomena are restricted to permutation maps, or perhaps they also
appear for genuinely quantum mechanical DU gates. The short recurrence times are most likely restricted to the circuits
that have close relations with the linear maps (over fields or rings). On the other hand, circuits with only multi-site
gliders could exist outside the realm of permutation maps, although it seems difficult to find other examples.

The reader might wonder, why did we restrict ourselves to DU gates, and why don't we perform similar searches among
models with arbitrary permutation maps. There were two motivations for focusing only on the DU circuits: On the one hand
side, in these models all conserved charges are gliders, and this makes the search for the charges especially simple. On the
other hand, these models (and their generalizations to multi-leg tensors, perfect or imperfect) are relevant for the
study of quantum scrambling and also for holography. Therefore, information about these circuits could be useful for
other studies as well.

The linear maps that we treated appear to be at a strange intersection of integrability and chaos. The general idea in
physics is
that linearity implies solvability, which is a counterpart of chaos. However, in these models some indicators of quantum
chaos signal complete ergodicity, at least for the local quantities. The behaviour is very similar to the discrete
version of Arnold's cat map, see \cite{cat-map}.

In these linear models we have linearity in the
classical variables over finite fields or rings. Therefore, the mathematical solvability is much more delicate than
usually and it has some surprises  (see the Appendix \ref{sec:app}). Linear maps over finite fields or rings
were discussed earlier (see for example \cite{set-th-YB-solutions,boll-classification}), but  we did not find in the
literature a proper solution of the 
circuits considered in this work. We believe it would be worthwhile to consider the linear models in more detail,
including those which involve rings but not fields.

Furthermore, it would be interesting to study the entanglement growth in the DU permutation models, in relation with the
number of gliders 
and the growth of the orbit lengths. Also, it would be important to find exact solutions for quenches in the
special models we considered, in analogy with the earlier work \cite{dual-unitary-4}. Such studies could highlight the
role of longer 
gliders in the emerging steady states. What type of Generalized Gibbs Ensemble do these models equilibrate to? This is
an interesting open question at the moment. 

In the present work we only focused on dual unitary circuits, but much of this material could be applied also to the
tri-unitary circuits introduced in \cite{triunitary}. Both the constructions for the gates, and the studies of the ergodicity could be
applied to these circuits. The integrability properties of tri-unitary gates have not yet been investigated, and it
would be interesting to consider Yang-Baxter structures also for those models.

\vspace{2cm}

\begin{acknowledgments}
The Appendix is a contribution from Roland Bacher (Universit\'e Grenoble Alpes, Institut Fourier, 100, Rue des
Math\'ematiques 38610 Gi\`eres, France, {\tt Roland.Bacher@univ-grenoble-alpes.fr}) and Denis Serre (\'Ecole Normale
Sup\'erieure de Lyon, U.M.P.A., UMR CNRS--ENSL 5669. 46 All\'ee d'Italie, 69364 Lyon cedex 07. France, {\tt
  denis.serre@ens-lyon.fr}).  We are thankful to Srinivasamurthy Aravinda, J\'anos Asb\'oth, Erika B\'erczi-Kov\'acs, Bruno Bertini,
Tam\'as Gombor, Arul Lakshminarayan, Katja Klobas, 
  Toma{\v z} Prosen, Lilla T\'othmer\'esz,  Ian Wanless and   Zolt\'an Zimbor\'as for useful discussions. 
\end{acknowledgments}

\appendix

\section{The recurrence time of the linear cellular automata}

\label{sec:app}

In this Appendix we treat the question posed in Section \ref{sec:perfect}: What is the recurrence time in the circuit
given by the perfect map \eqref{linp}? This is equivalent to finding the order of the matrix $\VV$ \eqref{VVdef} modulo
$p$. The problem requires the use of Fourier transform over finite fields.

For a reason that will become clear soon after, we use the standard notation $\F_p=\Z/p\Z$ for the field with $p$
elements. Let ${\bf M}_L(\F_p)$ stand for the ring of matrices of size $L\times L$ with coefficients in $\F_p$. Linear
algebra in ${\bf M}_L(\F_p)$ follows essentially the same rules as in ${\bf M}_L(\R)$ or ${\bf 
  M}_L(\C)$. Mind however that $\C$ is algebraically closed (every polynomial $P\in\C[X]$ splits over $\C$) but finite
fields are not.  

\subsection{Finite fields}

In a finite field $K$, the sum $1+\cdots+1$ vanishes when the number of summands is a
multiple of some number $p$, a prime number called the {\em characteristic}. For all $a\in K$, $pa=a+\cdots+a=0$. The
field can be viewed as a vector space over $\F_p$. Denoting its dimension $n$, its cardinal is thus $p^n$. A sound fact
is that, given $n\ge1$, there exists one and only one field with $q=p^n$ elements, up to isomorphism. Denoted $\F_q$, it
can be seen as the splitting field of the polynomial $X^q-X$: it is the smallest field containing $\F_p$ over which
the given polynomial completely factorizes.
Its multiplicative group $\F_q^\times$ is thus the set of
$(q-1)$th roots of unity. Contrary to the complex case, roots of unity cannot be expressed as exponentials, since an
exponential function cannot be defined in characteristic $p$. 

When $q=p^n$ and $q'=p^m$, $\F_q$ is contained in $\F_{q'}$ if and only if $n|m$ ($n$ divides $m$)~; we say that $\F_{q'}$ is an {\em extension} of $\F_q$, of degree $m/n$. More generally $\F_q\cap\F_{q'}=\F_r$ where $r=p^{{\rm gcd}(n,m)}$ and $\F_q\cdot\F_{q'}=\F_s$ where $s=p^{{\rm lcm}(n,m)}$.

Let $\ell\ge2$ be an integer. Contrary to the complex case, there do not always exist $\ell$ distinct $\ell$th roots
of unity. For instance, $1$ is the only $p$th root of unity in every extension of $\F_p$! But if $p\nmid\ell$ ($p$
does not divide $\ell$), then 
there exist $\ell$ distinct $\ell$th roots of unity. They belong to $\F_q$ where $q=p^n$ and $n$ is the smallest
non-zero number such that $\ell|p^n-1$. In
other words $p^n\equiv1$ mod $\ell$~; $n$ is the order of $p$ in the group $(\Z/\ell\Z)^\times$ of invertible elements
of $\Z/\ell\Z$. These roots form a multiplicative group denoted $U_\ell$.

\subsection{Fourier analysis of $\cal V$}

Let $L=2\ell$ be an even integer and $p$ be an odd prime number. We form the matrix ${\cal V}=AB$ as given by
\eqref{VVdef} and the problem is to evaluate the order $o_p(L)$ of ${\cal V}$ in ${\bf GL}_L(\F_p)$. For this, we
decompose vectors $x\in F^L$ into an odd and an even part: $y=(x_1,x_3,\ldots,x_{L-1})$ and $z=(x_2,x_4,\ldots,x_L)$,
both in $\F^\ell$. The images $x'=Ax$ and $x''=Bx$ are given by  
$$y_j'=y_j+z_j,\quad z_j'=y_j-z_j,$$
and
$$y_j''=-y_j+z_{j-1},\quad z_j''=y_{j+1}+z_j,$$
where the indices are understood mod $\ell$. The vector $X={\cal V}x$ is thus obtained through
\begin{equation}
\label{eq:YZyz}
Y_j=y_{j-1}-y_j-z_{j-1}-z_j,\qquad Z_j=y_j+y_{j+1}-z_j+z_{j+1}.
\end{equation}

Let $C$ be the cyclic shift operator acting on $L$ variables. The matrix $\VV$ commutes
with $C^2$: it is invariant with respect to translations of two sites.

Now our idea is to perform a Fourier transformation, to make use of this translational invariance. The standard step
would be to construct the eigenspaces of $C^2$ and to diagonalize the matrix $\VV$ within these eigenspaces. 
However, as opposed to the case of the complex numbers, the Fourier transform can not always be performed.
Sometimes there are not enough distinct eigenvalues of $C^2$ (which are roots of unity), furthermore, $C^2$
is not always diagonalizable. 

We set $L=2l$ and  work over a field $\F$, and also use the notation $C_\ell=C^2$.
The matrix
$\VV$ operates on the direct sum $\F[C_\ell]\oplus \F[C_\ell]$
of two copies of the group algebra $\F[C_\ell]$ of
the cyclic group $\mathbb Z/\ell\mathbb Z$; the two copies correspond to the odd and even parts of $c\in \F^L$
introduced above. The action of $\VV$ commutes with the obvious diagonal
action of $C_\ell$ on both copies (corresponding to odd
and even indices).

Observe that $\F[G_1\times G_2]=\F[G_1]\otimes_{\F} \F[G_2]$.
The order of $\VV$ over a finite field $\F=\mathbb \F_p$
is thus (up to a small factor given by a divisor of $p-1$) the least common
multiple of the order of $\VV(2a)$ and $\VV(2b)$ over $k$ where
$\ell=ab$ with $a$ and $b$ coprime.
For understanding the order of $\VV(L)$ over an odd prime $p$
it is thus essentially enough to understand the cases
$L=2p^m$ and $L$ coprime to $p$.

Below we will treat the two cases separately. The general strategy is the same in both cases: Understanding the
invariant subspaces for the action of $C^2$, and diagonalizing $M$ within these subspaces. However, the details of the
computations will be different.

\subsection{The case $L=2p^m$}

We consider now the case $\ell=p^m$ over the finite primary field
$\mathbb F_p$ for an odd prime $p$.
We have $\mathbb F_p[\mathbb Z/p^m\mathbb Z]=\mathbb F_p[t]/(1-t^{p^m})
=\mathbb F_p[t]/(1-t)^{p^m}$. A generator of the cyclic group
$\mathbb Z/p^m\mathbb Z$ acts thus on $\mathbb F_p[\mathbb Z/p^m\mathbb Z]$
by a unique Jordan-block $J$ of eigenvalue $1$ and maximal size $p^m$.
We denote by $T=J-\mathrm{Id}$ the associated nilpotent linear operator.
The obvious actions of $T\oplus T$ and $\VV$ on
$\mathbb F_p[\mathbb Z/p^m\mathbb Z]\oplus \mathbb F_p[\mathbb 
Z/p^m\mathbb Z]$
commute. The kernel of $T\oplus T$ has dimension $2$ and it consists of those vectors where both the odd
and even parts are constants.
Diagonalization of $\VV$ within this sub-space gives two eigenvectors with eigenvalues $\lambda_\pm=\pm 2\sqrt{-1}$,
which are elements of $\F_p$ or its extension $\F_{p^2}$, solving the equation $\lambda^2+4=0$.

Let $v$ be an arbitrary eigenvector of $\VV$. A suitable power of the
nilpotent operator $T\oplus T$ sends $v$ onto a non-zero element of
$\ker(T\oplus T)$. This implies that $v$ is also of eigenvalue
$\pm 2\sqrt{-1}$. The matrix $\VV^2$ is thus a union of Jordan-blocks
with common eigenvalue $-4$. Since we are working in characteristic $p$,
the matrix $\VV^{2p^m}$ is a non-zero multiple of the identity.

\begin{cor}
The order of $\VV$ over the finite field $\mathbb F_p$ equals thus
$ap^{\mu}$ for some $\mu\leq m$
where $a$ is the twice the order of $-4$ in the multiplicative group
$(\mathbb Z/p\mathbb Z)^\times$.
\end{cor}

We have thus proven that for the values $L=2p^m$ the order of $\VV$ grows at most linearly with $L$. This is a new result
of this Appendix; the linear growth was stated in \cite{chaos-qchann}, but a mathematical proof was not provided there. 

The exponent $\mu$ occuring in
the order is in $\{m-2,m-1,m\}$ since the matrix $\VV(L)$ is ``concentrated'' near the identity. More precisely,
$\VV(L)e_i$ is a linear combination with coefficients in $\{0,\pm 1\}$
involving only
$e_{i-2},\ldots,e_{i+2}$ (with indices read modulo $p^m$).
Moreover, $e_{i-1}$ and $e_{i+1}$ are involved with non-zero
coefficients. This implies that $\VV(L)$ has order at least $L/4$
over any finite field. We have thus $2(p-1)p^\mu\geq p^m/2$
implying $\mu\geq m-2$.

\subsection{The case $\ell$ co-prime to $p$}

Let us now consider the case when $\ell$ is not divisible by $p$. In such a case there exist enough roots of unities,
and the shift operator $C_\ell$ is diagonalizable. Therefore, this case is closer in spirit to the usual Fourier
transfrom known from linear systems over the complex numbers. 

Given an $\ell$th root of unity $\omega$, we define a vector $v_\omega=(1,\omega,\ldots,\omega^{\ell-1})$. If $U_\ell\subset\F_q$, then the vectors $v_\omega$, as $\omega\in U_\ell$, form a basis $\cal B$ of $(\F_q)^\ell$. The change of coordinates from the canonical basis to $\cal B$ is the discrete Fourier transform. We have the decomposition $(\F_q)^L=\oplus_{\omega\in U_\ell}E_\omega$ where $E_\omega$ is the plane defined by $y,z\parallel v_\omega$. If $y=av_\omega$ and $z=bv_\omega$, then (\ref{eq:YZyz}) yields $Y=\alpha v_\omega$ and $Z=\beta v_\omega$ with 
$$\binom\alpha\beta={\cal V}(\omega)\binom ab,\qquad{\cal V}(\omega)=\begin{pmatrix} \omega^{-1}-1 & -\omega^{-1}-1 \\ \omega+1 & \omega-1 \end{pmatrix}.$$
The matrix $\cal V$ is thus conjugated to the block-diagonal matrix ${\cal V}'$ whose diagonal blocks are the ${\cal
  V}(\omega)$'s. In particular, its order is the lcm of the orders of the blocks in ${\bf GL}_2(\F)$. Remark that  ${\cal
  V}(\omega)$ and  ${\cal V}(\omega^{-1})$ are conjugated to each other.

The characteristic polynomial of ${\cal V}(\omega)$ is $\lambda^2+(2-\omega-\omega^{-1})\lambda+4$. It admits disctinct roots provided $\omega\ne-1$ and $\omega+\omega^{-1}\ne6$. If so,  ${\cal V}(\omega)$ is diagonalizable and its order is the lcm of the order of its eigenvalues. The order of  ${\cal V}(-1)=-2I_2$ is just the order of $-2$ in $\F_p$. If $\omega+\omega^{-1}=6$, then ${\cal V}(\omega)$ is not diagonalizable and its double eigenvalue is $\lambda=2$~; its order is $p$ times the order of $2$ in $\F_p$.

Denote $n$ the order of $p$ mod $\ell$, and $q=p^n$ so that $U_\ell\subset\F_q$. One can prove that $\omega+\omega^{-1}$ belongs to $\F_{p^s}$ with $s=n$ if $n$ is odd, or $s=n/2$ if $n$ is even. Since the eigenvalues of ${\cal V}(\omega)$ obey a quadratic equation with coefficients in $\F_{p^s}$, they belong to $\F_{p^{2s}}$. In particular their order divides $p^{2s}-1$. We conclude that $o_p(L)$ divides ${\rm lcm}(p(p-1),p^{2s}-1)$, hence
\begin{prop}\label{p:div}
The order of $\cal V$ divides $p(p^{2s}-1)$, hence $p(p^{2n}-1)$.
\end{prop}
Since $n$ divides the order $\phi(\ell)$ of the group $(\Z/\ell\Z)^\times$ ($\phi$ is Euler totient function) and $\phi(\ell)\le\ell-1$ (with equality if $\ell$ is prime), we infer the exponential bound
\begin{cor}
We have $o_p(L)\le p^{L-1}-p$.
\end{cor}

With this we have obtained a rigorous upper bound for the exponential growth.  

\subsection{The case of $p$-repunits: quadratic growth.}

The exponential bound is much too big in some cases. For example let us consider the case where $\ell=p_n:=\frac{p^n-1}{p-1}$\, (called a $p$-repunit because it writes $1\cdots1$ in base $p$). The letter $n$ is chosen on purpose: it is the order of $p$ mod $\ell$. Then Proposition (\ref{p:div}) gives a quadratic bound
$$o_p(L)\le p\left[((p-1)\ell+1)^2-1\right]=p(p-1)\ell((p-1)\ell+1).$$

This subset of volumes with quadratic growth was not noticed in \cite{chaos-qchann}.

\providecommand{\href}[2]{#2}\begingroup\raggedright\endgroup

\end{document}